\documentclass[%
superscriptaddress,
amsmath,amssymb,
twocolumn,
aps,pra,
floatfix,
]{revtex4-2}

\usepackage{graphicx}
\usepackage{dcolumn}
\usepackage{bm}
\usepackage{color}
\usepackage{qcircuit}
\usepackage{booktabs,eqparbox}
\usepackage{braket}
\usepackage{dsfont}
\usepackage{placeins}
\usepackage{comment}
\usepackage[utf8]{inputenc}
\usepackage{pgfplots}
\usepackage{multirow}
\usepackage{amsmath}
\usepackage{makecell}
\usepackage{float}
\usepackage{cellspace}
\usepackage{tabularx}
\usepackage{lipsum}
\DeclareUnicodeCharacter{2212}{−}
\usepgfplotslibrary{groupplots,dateplot}
\usetikzlibrary{patterns,shapes.arrows}
\pgfplotsset{compat=newest}

\usepackage{hyperref}
\usepackage{cleveref}
\newcommand\numberthis{\addtocounter{equation}{1}\tag{\theequation}}

\DeclareMathOperator\Tr{Tr}

\begin{document}
\title{Optimizing Superconducting Three-Qubit Gates for Surface-Code Error Correction}

\author{Stephan Tasler}
\email{stephan.tasler@fau.de}
\affiliation{Physics Department, Friedrich-Alexander-Universität Erlangen Nürnberg, Germany}
\author{Josias Old}
\email{j.old@fz-juelich.de}
\affiliation{Institute for Quantum Information, RWTH Aachen University, D-52056 Aachen, Germany}
\affiliation{Peter Grünberg Institute, Theoretical Nanoelectronics, Forschungszentrum Jülich, D-52425 Jülich, Germany}

\author{Lukas Heunisch}
\affiliation{Physics Department, Friedrich-Alexander-Universität Erlangen Nürnberg, Germany}
\author{Verena Feulner}
\affiliation{Physics Department, Friedrich-Alexander-Universität Erlangen Nürnberg, Germany}
\author{Timo Eckstein}
\affiliation{Physics Department, Friedrich-Alexander-Universität Erlangen Nürnberg, Germany}
\author{Markus M\"uller}
\affiliation{Institute for Quantum Information, RWTH Aachen University, D-52056 Aachen, Germany}
\affiliation{Peter Grünberg Institute, Theoretical Nanoelectronics, Forschungszentrum Jülich, D-52425 Jülich, Germany}
\author{Michael J. Hartmann}
\affiliation{Physics Department, Friedrich-Alexander-Universität Erlangen Nürnberg, Germany}
\date{\today}

\begin{abstract}
Quantum error correction (QEC) is one of the crucial building blocks for developing quantum computers that have significant potential for reaching a quantum advantage in applications. Prominent candidates for QEC are stabilizer codes for which periodic readout of stabilizer operators is typically implemented via successive two-qubit entangling gates, and is repeated many times during a computation. 
To improve QEC performance, it is thus beneficial to make the stabilizer readout faster and less prone to fault-tolerance-breaking errors.
Here we design a 3-qubit CZZ gate for superconducting transmon qubits that maps the parity of two data qubits onto one measurement qubit in a single step. We find that the gate can be executed in a duration of $35\,$ns with a fidelity of F$=99.96 \, \%$. To optimize the gate, we use an error model obtained from the microscopic gate simulation to systematically suppress Pauli errors that are particularly harmful to the QEC protocol. Using this error model, we investigate the implementation of this 3-qubit gate in a surface code syndrome readout schedule. We find that for the rotated surface code, the implementation of CZZ gates increases the error threshold by nearly 50\% to $\approx 1.2\,\%$ and decreases the logical error rate, in the experimental relevant regime, by up to one order of magnitude, compared to the standard CZ readout protocol. We also show that for the unrotated surface code, strictly fault-tolerant readout schedules can be found. This opens a new perspective for below-threshold surface-code error correction, where it can be advantageous to use multi-qubit gates instead of two-qubit gates to obtain a better QEC performance.

\end{abstract}
\maketitle

\section{Introduction}

Quantum computing is regarded as a technology that enables solving particular computationally hard problems much faster than classical computation\,\cite{Dalzell2025}. For this reason the interest in building universal quantum computers has experienced a rapid growth over the last decades. Although the fidelities of physical quantum gates have steadily improved over the last years, quantum circuits are still vulnerable to residual errors. Therefore quantum error correction (QEC) is vital in developing scalable, fault-tolerant quantum computers.

Promising candidates for realizable near-term QEC codes are originating from the group of topological error correcting codes\,\cite{Kitaev2003,Kitaev1997}. In this work we focus on the surface code\,\cite{Bravyi1998,Dennis2002,Fowler2012}, which is a particularly suitable topological code for the implementation on superconducting circuit hardware, as it can be implemented on a 2D lattice chip, using only nearest-neighbour interactions. It also has one of the highest circuit-level noise error thresholds of about $1 \%$\,\cite{Wang2011}. Recent experiments demonstrated increasingly larger implementations on superconducting hardware\,\cite{Andersen2020,Krinner2022,Acharya2023,Acharya2024}, including first entangling operations between logical qubits via lattice surgery\,\cite{Besedin2025}.

The surface code is a stabilizer code, where repetitive measurements of stabilizers are executed during the error correction process. Stabilizer operators don't change the encoded logical quantum state but, if measured, give information about errors that may have occurred. Successful QEC typically requires frequent stabilizer measurements, which, in the surface code, are commonly done via a sequence of two-qubit gates.

Given the high number of repeated error correction cycles during a quantum computation and the limitation of correctable physical qubit errors imposed by the fidelity of the two-qubit gates\,\cite{Fowler2012}, improving the stabilizer measurement protocol can significantly accelerate the overall computation and boost QEC performance. By parallelizing the stabilizer readout \,\cite{DiVincenzo2013,Ciani2017}, the duration of a stabilizer measurement cycle can be reduced. Several studies for different platforms have shown that parallelized stabilizer readout protocols can increase the physical error correction threshold of surface code QEC schedules\,\cite{Schwerdt2022,Uestuen2024,Reagor2022}. Moreover, gate optimizations for suppression of experimental systematic errors\,\cite{Cerfontaine2020} and for logical qubit performance\,\cite{Jandura2023} in the context of erasure conversion in Rydberg atoms have been proposed.

In this work, we design and optimize a hardware-specific gate for parallelizing stabilizer readout in superconducting qubits and show that the best performance of the error correcting code is not achieved when maximizing the fidelity of the gate, as common knowledge would suggest. Instead, the performance of the code is further enhanced, if, among all errors that are generated by the gate, those, which result in errors that cannot be corrected by the surface code of a given distance, are suppressed the most. 
\begin{figure} [h!]
    \centering
    \includegraphics[width=\linewidth]{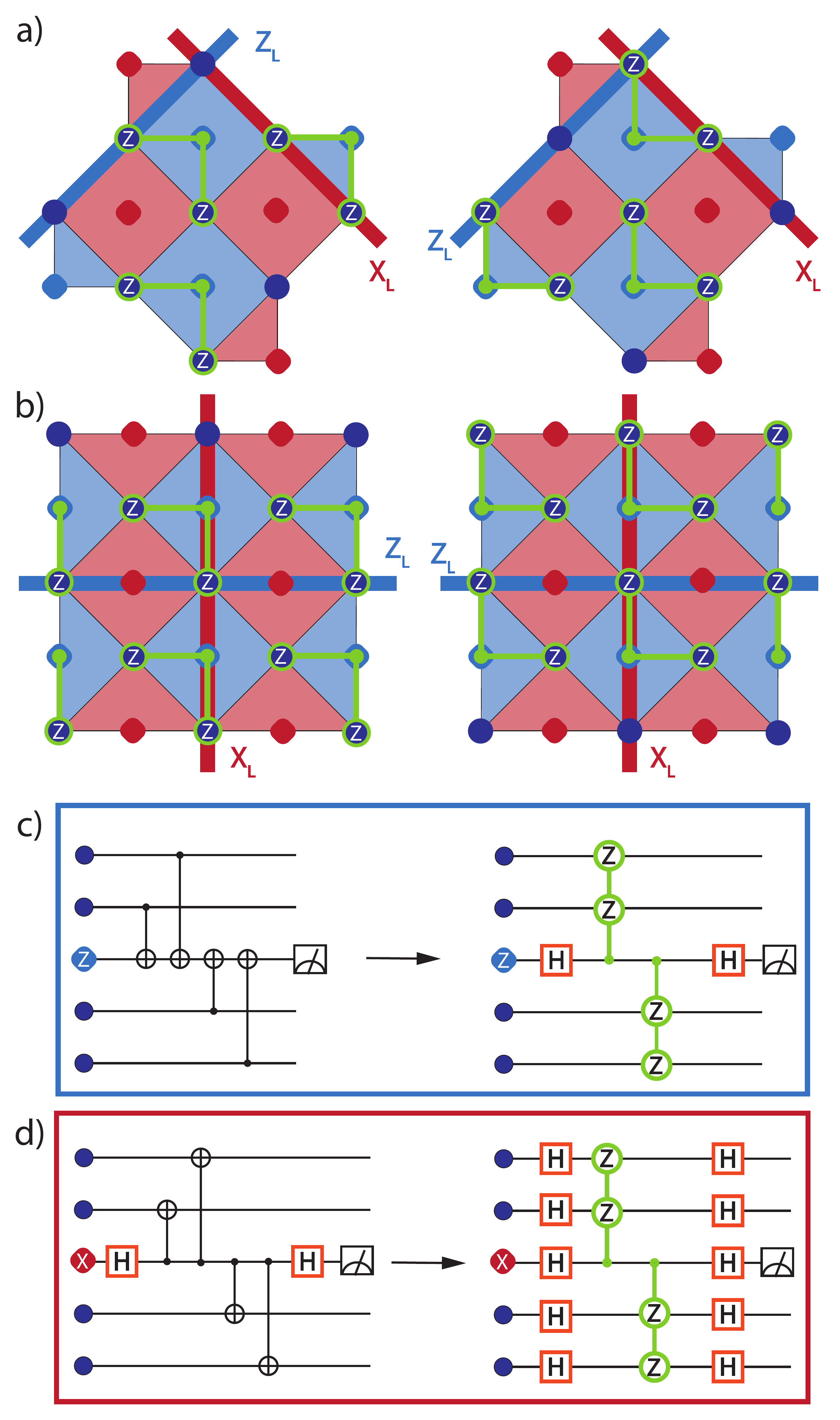}
    \caption{Schematic illustration of the unrotated (a) (rotated (b)) surface code readout schedule. The dark blue circles represent the data qubits, the blue (red) squares show the $Z$- ($X$-) stabilizer plaquettes. The here depicted $Z$-stabilizer readout consists of two steps: in each step multiple CZZ gates represented by the green lines are applied in parallel. The thick blue (red) lines depict the logical $Z$- ($X$-) Pauli operators of the underlying surface code. Because of the larger qubit overhead, the unrotated surface code has a higher robustness against fault-tolerance-breaking errors, which allows for a fault-tolerant CZZ readout schedule. c) shows the readout circuits for the $Z$- ($X$-) stabilizer displayed in the blue (red) box. The commonly used readout circuit, based on four sequential two-qubit gates, is displayed on the left side, the readout using the CZZ gates as proposed in this work is depicted on the right side.}
    \label{fig:surface_stabilizer_readout}
\end{figure}
More specifically, we parallelize the stabilizer readout protocol, as depicted in Fig.\,\ref{fig:surface_stabilizer_readout} c) and d), by designing a CZZ gate for superconducting qubits, where one qubit acts as a measurement qubit and the other two qubits act as data qubits. The CZZ gate can be viewed as a $Z$ operation that is applied on the measurement qubit, controlled by the parity of the two data qubits \footnote{Note that the 'control' of the CZZ gate is on the measurement qubit, but since $\mathrm{CZZ} = \ket{0}\bra{0} \otimes II + \ket{1}\bra{1} \otimes ZZ = I \otimes (\ket{00}\bra{00} +\ket{11}\bra{11}) + Z \otimes (\ket{01}\bra{01} +\ket{10}\bra{10})$, this gate can also be thought of as a $X$-parity-controlled $Z$-gate.}. This CZZ gate can thus be used as a parity mapping gate for the $Z$-, $X$- and $Y$-stabilizer readout, which makes it a natural building block of two-qubit stabilizer readouts as they are needed in sub-system QEC codes\,\cite{Kribs2005,Kribs2005a} and Floquet codes\,\cite{Hastings2021, Davydova2023, MagdalenadelaFuente2025}. In addition to this, it is well-adopted for surface-code error correction, where only $Z$-stabilizer measurements (Fig.\,\ref{fig:surface_stabilizer_readout} c)) and $X$-stabilizer measurements (Fig\,\ref{fig:surface_stabilizer_readout} d)) are needed. We optimize this to maximize the logical qubit fidelity by explicitly analyzing which Pauli errors it causes and aiming to suppress those Pauli errors which harm the surface code QEC procedure most.

We find that, by using the QEC-performance-optimized CZZ gate, the logical error of the rotated surface code can be suppressed by about one order of magnitude for higher code distances in the range of experimentally realistic physical qubit error rates. Moreover, the physical error threshold can be increased significantly, by almost 50\%, from $\approx0.66\,\%$ to $\approx1.1\,\%$. For the unrotated surface code we find a similar physical error threshold improvement from $\approx 0.66\,\%$ to $\approx 1.1\,\%$. Although the unrotated surface code is operated in a fault-tolerant readout schedule, we still find an improvement in the logical error rate.

The paper is organized as follows: Section \ref{sec:parallelized_cz_gate} introduces the fundamental concept of the CZZ gate under investigation and discusses the proposed parallelized parity-check protocol. In Section \ref{sec:circuit_analysis}, we provide a detailed quantum electrodynamic description of the CZZ gate. Section \ref{sec:gate_simulation} covers the derivation of the gate-based Pauli error model, and addresses the suppression of fault-tolerance-breaking errors using optimal control. Finally, we discuss the performance of the resulting surface-code error correction protocol in Section \ref{sec:application_on_the_surface_code_qec_protocol}.

\section{Parallelized CZ gates} \label{sec:parallelized_cz_gate}
\begin{figure}
    \centering
    \includegraphics[width=0.50\textwidth]{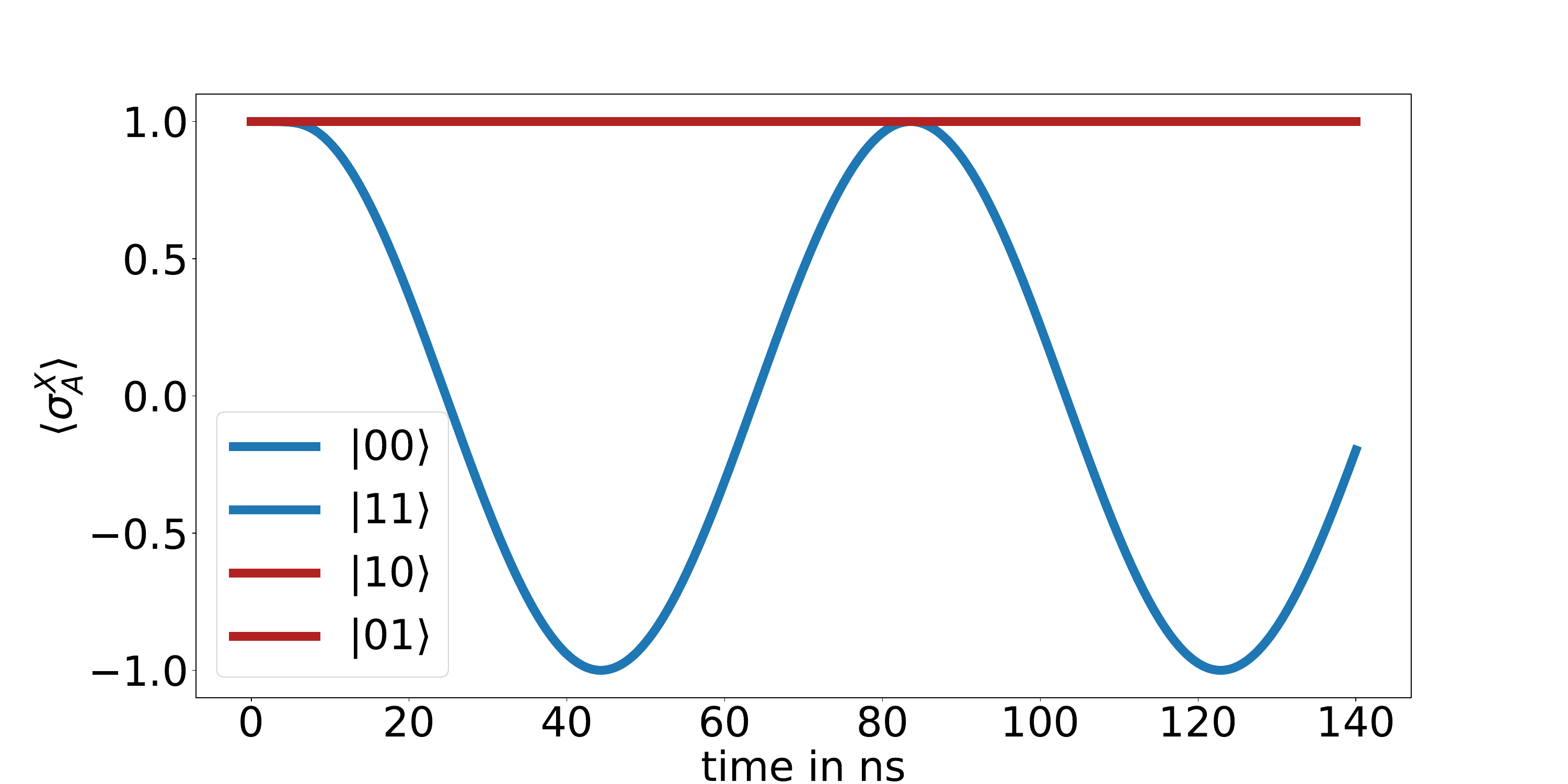}
    \caption{Dynamics for the idealized toy model of Eq.\,(\ref{eqn:ideal_ham}). Depending on the parity of the states of the outer qubits, the center qubit oscillates between the states $\ket{+}$ and $\ket{-}$. Depending on the value of $J$, here at $45\,$ns, there is a maximal contrast between the oscillations corresponding to even and odd $ZZ$-parity input states.}
    \label{toy_model}
\end{figure}

In this work we focus on the execution of two parallelized controlled-$Z$ (CZ) gates forming a CZZ gate.
CZ gates are generated by $ZZ$-interactions. Hence, the fundamental concept of parallelizing such CZ gates is based on the commutativity of the involved $ZZ$-interactions, which allows one to combine the execution of multiple CZ gates in one step. 
For an optimal parallelization of the CZ gates for three qubits (labeled Q$_1$,Q$_2$,Q$_3$), the $ZZ$-interactions between the two outer qubits (Q$_1$,Q$_3$) and the center qubit (Q$_2$) should be identical. We can describe the ideal interaction by a simplified two-level Hamiltonian,
\begin{equation} \label{eqn:ideal_ham}
    H = \sum_{i=1,3} J_i \sigma_i^Z \sigma_2^Z \, ,
\end{equation}
where $J_i$ denotes the coupling strength, $\sigma_i^Z$ is the Pauli-$Z$ matrix of the $i$-th (outer) qubit and $\sigma_2^Z$ the Pauli-$Z$ matrix of the 2nd (center) qubit. The dynamics generated by this Hamiltonian is shown in Fig.~\ref{toy_model}. 
The parallel $ZZ$-interactions cause a phase oscillation, which is dependent on the parity of the two outer qubits. We choose the duration of the $ZZ$-interactions in a way that the phase accumulation is equal to the one obtained by applying a CZ gate between each of the $i$-th (outer) qubit and the 2nd (center) qubit. The resulting dynamics then corresponds to the unitary
\begin{equation}
    U_{\text{CZZ}} = \begin{pmatrix}
    1 & 0 & 0 & 0 & 0 & 0 & 0 & 0 \\
    0 & 1 & 0 & 0 & 0 & 0 & 0 & 0 \\
    0 & 0 & 1 & 0 & 0 & 0 & 0 & 0 \\
    0 & 0 & 0 & -1 & 0 & 0 & 0 & 0 \\
    0 & 0 & 0 & 0 & 1 & 0 & 0 & 0 \\
    0 & 0 & 0 & 0 & 0 & 1 & 0 & 0 \\
    0 & 0 & 0 & 0 & 0 & 0 & -1 & 0 \\
    0 & 0 & 0 & 0 & 0 & 0 & 0 & 1 \\
    \end{pmatrix} \, , \label{eqn:ideal_matrix}
\end{equation}
with matrix elements $(U_{\text{CZZ}})_{j,l} = \bra{\tilde{\text{Q}}_1,\tilde{\text{Q}}_2,\tilde{\text{Q}}_3} U_{\text{CCZ}}\ket{\text{Q}_1,\text{Q}_2,\text{Q}_3}$, where $\tilde{\text{Q}}_1,\tilde{\text{Q}}_2$ and $\tilde{\text{Q}}_3$ (Q$_1$,Q$_2$ and Q$_3$) are the digits of $j =0, 1, \dots 7$ ($l =0, 1, \dots 7$) in binary representation.
As our aim is to use this gate for parity checks, we now discuss a protocol that maps the parity of the two outer qubits (Q$_1$, Q$_3$) onto the state of the center qubit (Q$_2$).

\subsection{Parity Check Protocol}
\begin{figure}
    \centering
    \includegraphics[width=\linewidth]{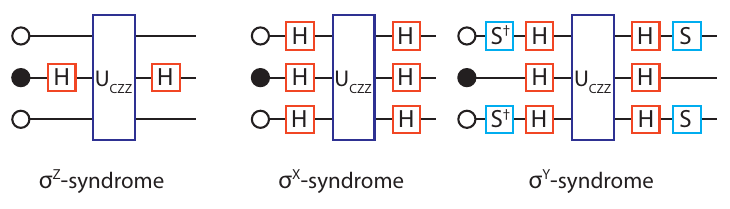}
    \caption{The CZZ gate can be used to measure the $Z$-,$X$- and $Y$-stabilizer, by parallelizing the $ZZ$ interactions between the two outer qubits (circle with white surface) and the middle qubit (circle with black surface).}
    \label{fig:all_syndroms_XYZ}
\end{figure}

For the following discussion of the parity check measurement procedure we focus only on checks of $Z$-parity. $X$- and $Y$-parity checks can be easily obtained by adding additional single qubit rotations on each qubit at the beginning and the end of the circuit. To simplify the reading of this paper, we will refer to Hadamard and S gates when mentioning the basis rotations for the different stabilizers. It should be noted, however, that in an experiment, both the Hadamard and S gates are executed by single-qubit rotations $\left(\text{R}_X(\theta), \text{R}_Y(\theta), \text{R}_Z(\theta) \right)$. The readout protocols for the individual  bases are shown in Fig.~\ref{fig:all_syndroms_XYZ}.

For the further discussion, we divide the protocol into five steps. In the first step, we prepare the center (or measurement) qubit in the $X$-basis eigenstate $\ket{+}$ with a Hadamard gate. In the second step we activate multiple parallel $ZZ$-interactions between the center qubit and the outer qubits. This causes an oscillation of the center qubit between the $\ket{+}$ and $\ket{-}$ states. Thus, if the $ZZ$-interactions have the same strength, the oscillation depends on the parity of the data qubits, and after a time that depends on the strength of the $ZZ$-interactions, the Pauli $X$ expectation value of the ancilla qubit is perfectly split depending on whether the outer qubits have even or odd parity, see Fig.\,\ref{toy_model}. At this point, in a third step, we switch off the $ZZ$-interaction and apply in a fourth step a second Hadamard gate to the center qubit. This transforms the center qubit back into the $Z$-basis eigenstate ($\ket{+}\to\ket{0}$ and $\ket{-}\to\ket{1}$). The final step involves measuring the measurement qubit to extract the parity of the outer qubits.

\section{Circuit analysis} \label{sec:circuit_analysis}
\begin{figure}
    \centering
    \includegraphics[width = 0.5\textwidth]{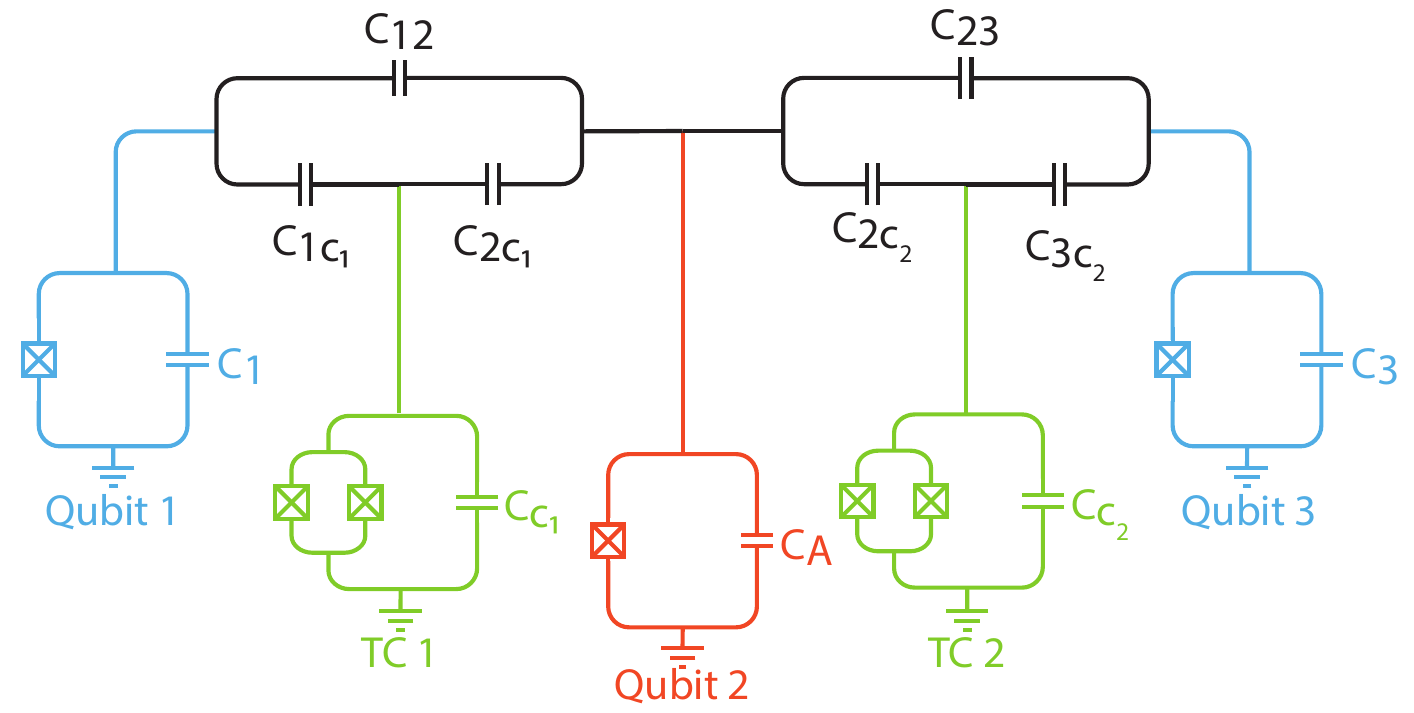}
    \caption{The parallel CZ gate is implemented on a superconducting circuit, that consists of three fixed-frequency transmons (blue, orange) and two frequency-tunable transmons (tunable couplers, green). By simultaneously tuning the two tunable couplers, parallel $ZZ$-interactions between the outer qubits and the center qubit can be engineered.}
    \label{circuit_diagram}
\end{figure}

While the simplified model explains the working principle, it does not describe the full dynamics of a possible experimental implementation. To investigate this in more depth, we consider a system consisting of three transmon qubits (Q$_1$,Q$_2$,Q$_3$) and two intermediate tunable couplers (TC$_1$, TC$_2$), displayed in Fig.\,\ref{circuit_diagram}, in a lumped-element description.
For transmon qubits, there are multiple techniques for designing a CZ gate. In a system of two qubits coupled by a coupler ($\ket{\text{qubit}, \text{coupler}, \text{qubit}}$, one technique is to bring the states $\ket{101}$ and $\ket{200}$ in resonance and control the interaction time to accumulate a desired phase in a Larmor precession \cite{Sung2021}.
Another approach is to use dispersive shifts, which alter specific eigenenergies of the system \cite{Collodo2020,Heunisch2023}. To engineer and tune the desired $ZZ$-interaction, these experiments employ a controllable coupling element, allowing them to regulate the gained phase via the duration of the interaction. This adiabatic approach to CZ gates offers the advantage of avoiding modifications to qubit transition frequencies, which can cause unwanted resonances with spectator qubits. We here adopt this second method to design a strategy for executing two CZ gates in parallel.

The circuit we investigate consists of five circuit elements, two of which are frequency-tunable couplers and three are fixed-frequency transmon qubits consisting of a Josephson junction and a large shunt capacitance. While fixed-frequency transmons tend to have longer coherence times, one could also consider frequency-tunable transmons for enhanced flexibility, e.g. to avoid resonances with two-level defects. The circuit diagram is shown in Fig.~\ref{circuit_diagram}. 
 In the transmon regime the qubits are well protected against charge noise. The two couplers (green) are frequency-tunable, i.e.~ they each consist of a capacitively shunted DC-SQUID \cite{Vool2017, Rasmussen2021}.
Our design keeps the fixed-frequency qubits at their sweet spot while only adjusting the frequency of the tunable couplers \cite{Collodo2020,Sung2021,Li2022}.

To analyze the physics of the circuit depicted in Fig.~\ref{circuit_diagram}, we derive the following Hamiltonian using canonical circuit quantization \cite{Vool2017,Rasmussen2021}, 
\begin{align*}
    H&= \sum_{i=1,2,3}\omega_ib_i^\dagger b_i +\frac{\alpha_i}{2}b_i^\dagger b_i^\dagger b_i b_i \\ 
    &+\sum_{j=1,2}\omega_{\text{c}_j}c_j^\dagger c_j +\frac{\alpha_j}{2}c_j^\dagger c_j^\dagger c_j c_j \\
    &+\sum_{k=1,2,3; l=1,2} g_{k,c_l}(b_k - b_k^\dagger) (c_l - c_l^\dagger)\\
    &+\sum_{m<o=1,2,3} g_{mo} (b_m - b_m^\dagger) (b_o - b_o^\dagger)\\
    &+ g_{c_1,c_2}(c_1 - c_1^\dagger) (c_2 - c_2^\dagger) \, . \numberthis \label{eqn:2QA_circuit_Hamiltonian}
\end{align*}
Here the $b_i^{\dagger}$ ($b_i$) are the bosonic creation (annihilation) operators for the three qubits ($i=1,2,3$) and 
$c_j^{\dagger}$ ($c_j$) are the creation (annihilation) operators for the two tunable couplers ($j=1,2$). For simplicity, the constant terms were neglected. The frequencies are denoted by $\omega_i$, the anharmonicities by $\alpha_i$ and the individual couplings by $g_{i,j}$, where the indices correspond to the respective qubits or couplers. 

The qubit frequencies are given in terms of charging and Josephson energies by $\omega_i = \sqrt{8E_{C_{ii}}E_{J_i}}-E_{C_{ii}}$ for $i=1,2,3$ (qubits) and $i=\text{c}_1,\text{c}_2$ (couplers). The anharmonicities are given by $\alpha_i = - E_{C_{ii}}$ for $i=1,\text{c}_1,2,\text{c}_2,3$. The qubit-couplings are expressed by $g_{k,c_l} = -4 E_{C_{kc_l}} (\sqrt{\zeta_k \zeta_{\text{c}_l}})^{-1}$, where $k=1,2,3 \ \text{and} \ l=1,2$ and the qubit-qubit couplings are represented by $g_{mo} = -4 E_{C_{mo}} (\sqrt{\zeta_m \zeta_o})^{-1}$, with $m<o=1,2,3$ and $g_{c_1,c_2} = -4 E_{C_{c_1c_2}} (\sqrt{\zeta_{\text{c}_1} \zeta_{\text{c}_2}})^{-1}$ defines the coupler-coupler coupling. In these expressions, the impedance is defined as $\zeta_i = \sqrt{(8E_{C_{ii}})/E_{J_i}}$ for $i=1,\text{c}_1,2,\text{c}_2,3$.
Here $E_{C_{xy}}$ describes the capacitative energy for the corresponding active circuit nodes $x,y$. $E_{J_x}$ represents the Josephson energy for the active circuit node $x$, where the active circuit nodes are defined as $x,y \in (1,\text{c}_1,2,\text{c}_2,3)$, for a more detailed circuit quantization see Appendix \ref{apx:circuit_quantisation}. 

To analyze the parallel execution of two CZ gates, we investigate tuning the couplers with two magnetic fluxes near the frequency of the qubits. This causes the prepared initial state to acquire a phase that depends on the state of the qubits. Besides dynamics in the qubit subspace, leakage into non-computational states will also occur during the execution of the gates. The control pulses of the couplers therefore have to be chosen in a way that, after the gate, any leakage population has returned into the computational subspace. 

In an experimental realization, the relevant computational basis is formed by the eigenstates of the Hamiltonian in Eq.\,\ref{eqn:2QA_circuit_Hamiltonian} in the idle regime, where interactions between the individual qubit-circuits are maximally suppressed. Since these computational basis states do not deviate strongly from the bare states $\ket{ijklm}_b$, where the indices $i,j,k,l,m$ correspond to the excitation numbers of $\text{Q}_1$,$\text{TC}_1$,$\text{Q}_2$,$\text{TC}_2$,$\text{Q}_3$, we use the same indices, for the computational basis state $\ket{ijklm}$ that has maximal overlap with the bare state $\ket{ijklm}_b$. 

\section{Gate optimization} \label{sec:gate_simulation}
\begin{figure}
    \centering
    \includegraphics[width=0.5\textwidth]{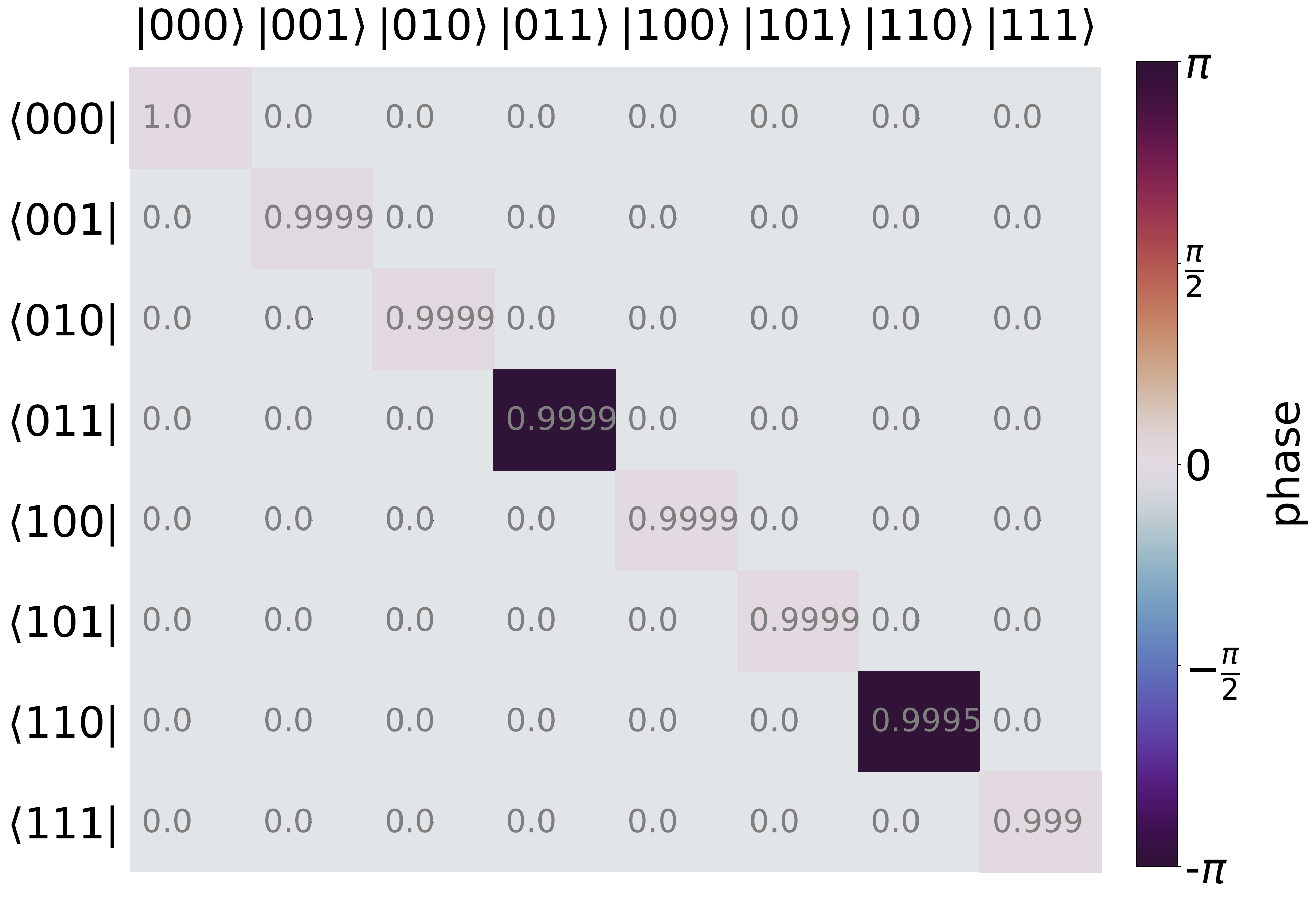}
    \caption{The simulated channel $\mathcal{U}_{\text{sim}}$ for the $35\,$ns gate is a very good approximation of the target unitary $U_{\text{ideal}}$, up to an irrelevant global phase. The relative phase between input state and the state after the parity check gate is depicted by the color of the squares. The size of the squares and the light grey number corresponds to the occupation probability of the input and time-evolved states. More detail about obtaining the simulated unitary matrix is given in App.\,\ref{apx:parameter_optimization}.}
    \label{parit_check_fidelity}
\end{figure}

For obtaining an optimal parity check gate, we determined adequate circuit and pulse parameters. Since the gate should involve the same $ZZ$-interaction between each data qubit and the ancilla qubit, it is beneficial to choose symmetric parameters for the circuit model given in Fig.\,\ref{circuit_diagram}. For the capacitances and Josephson energies of the fixed frequency qubits we choose an experimentally feasible value range \cite{Collodo2020, Baker2022}. 
We then optimize the control pulses, i.e. the flux pulses that are applied to the tunable couplers for this purpose. Details of the optimization procedure and final values for the fixed circuit parameters and control pulses are provided  in Appendix \ref{apx:parameter_optimization}.
In this way, we obtain gate pulses and unitaries $U_\text{sim}$(see Fig.~\ref{parit_check_fidelity} for an example) which are derived from the numerical time-evolution generated by the Hamiltonian given in Eq.\,\ref{eqn:2QA_circuit_Hamiltonian}. To quantify the gate performance, one commonly calculates the gate specific error as,
\begin{equation}\label{eq:epsilon_gate}
    \varepsilon_{\text{gate}} \approx 1 - \text{F} = 1 - \frac{\Bigl|\Tr\left( U_{\text{sim}}^\dagger U_{\text{CZZ}} \right)\Bigr|}{8} ,
\end{equation}
where $F$ is the fidelity of the gate. To make a more realistic assessment, we here also consider the error due to decoherence, which we estimate as
\begin{equation}
    \varepsilon_{\text{dec}} \approx 1- e^{-\frac{t_{\text{gate}}}{\tau}}.
\end{equation}
and assume a coherence time of a transmon to be $\tau \approx 50 \, \mu$s \cite{Wang2022,Sung2021}. We design our gate times such that the decoherence error and the gate error are of the same order of magnitude. Even though that the gate error could be reduced further at the expense of a longer gate time, this would not lead to a benefit as the accuracy of the gate would still be limited by decoherence. Considering a shorter gate, on the other hand would reduce the effects of decoherence but lead to larger gate errors. 

\subsection{Error model and error based optimization} \label{subsec:error_model}
The fidelities we obtain by simulating the CZZ gate give a good measure of the performance of a single CZZ gate. However, our main use case for the CZZ gate is to improve surface code quantum error correction. To assess the CZZ gate performance in a surface code error correction protocol, we therefore seek an error model for the gate in terms of Pauli errors. To do so, we take our process matrix displayed in Fig.~\ref{parit_check_fidelity} and follow Ref.~\cite{Perlin2023} to obtain an effective Pauli noise model for our gate. We write the error channel as a composition of the simulated channel $\mathcal{U}_{\text{sim}}$ and the ideal quantum channel $\mathcal{U}_\text{CZZ} = U_\text{CZZ} \otimes U_\text{CZZ}^*$. Introducing Pauli errors $P$ and $Q$ we write the closest Pauli error channel as
\begin{equation}
    \mathcal{U}_{\text{err}} = \mathcal{U}_{\text{sim}} \circ \mathcal{U}_\text{CZZ}^{-1} = \sum_{PQ} w_{PQ} P \otimes Q^*.   
\end{equation}
Here "closest" means with respect to the metric induced by the norm $|| A || = \sqrt{\braket{A,A}}$ with the normalized Frobenius inner product $\braket{A,B} = \Tr(A^\dagger B)/\text{dim}(A)$. The Pauli coefficients $w_{PQ}$ correspond to the probability of the Pauli channel $PQ$ and $\circ$ denotes the successive execution of two channels.
We calculate the Pauli coefficients according to \cite{Perlin2023} by flattening $\mathcal{U}_{\text{err}}\left( \ket{a}\bra{b}\right)$ into a column vector $|\mathcal{U}_{\text{err}},ab) = \mathcal{U}_{\text{err}}|ab)$. With this technique we calculate all the $4^N$ column vectors for all $a,b \in \{0,1\}^N$, to obtain the matrix $M = \sum_{a,b} \mathcal{U}_{\text{err}}|ab)(ab|$ and calculate the Pauli coefficients $w_{PQ}$ by taking the normalized Frobenius inner product $w_{PQ} = \braket{P \otimes Q^*, M}$.

We are only interested in the symmetric Pauli channels, which we define as $w_{PP}$. Therefore we apply the Pauli twirling approximation (PTA)\,\cite{Geller2013} to neglect the asymmetric Pauli coefficients. The asymmetric Pauli coefficients correspond to the leakage out of the computational subspace and are neglected by the PTA. We ensure that leakage is suppressed via an additional contribution to the cost function used in the optimization. In an experimental realization, remaining leakage errors can also be converted into Pauli errors via reset protocols\,\cite{McEwen2021, Miao2023,Jandura2024,Lacroix2025}. 
As the diagonal elements corresponding to the PTA do not sum to unity, we normalize their coefficients. We then define a base error probability $\tilde{p} = 1-w_{III}$, where $w_{III}$ denotes the Pauli coefficient for the Pauli channel $III$. In our surface code simulations where we vary a physical error parameter $p$, we thus rescale all Pauli coefficients such that $1-w_{III} = \sum_{P \neq III} w_P = p$.

A core feature of our gate optimization is the choice of our cost function that can address multiple goals depending on the task of the gate. In our case we want to design a gate, which specifically suppresses fault-tolerance-breaking errors in the surface code QEC. We design our cost to consist of three terms, which we weight by different weights $\eta_1, \eta_2$ and $\eta_3$. We can thus fine-tune our cost by adapting the weights $\eta_j$ according to our aims. Here we define our cost as
\begin{equation}
    \text{cost} = \eta_{1} \varepsilon_\text{L} + \eta_{2} \sum_{P} w_{PP} + \eta_{3} \sum_{K} \gamma_{K} \cdot w_{KK} . \label{eqn:new_cost} 
\end{equation}
Independently of the task of the gate, we want to minimize leakage out of the qubit subspace. This is ensured by the first term $\eta_{1}\varepsilon_\text{L} = \eta_{1}\left(1 - \frac{1}{8} \cdot \Bigl|\Tr\left( U_{\text{sim}}^\dagger U_{\text{sim}} \right)\Bigr|\right)$ that penalizes such leakage. The second term that is implemented in the cost regardless of the task is the sum over all unweighted Pauli coefficients $w_{PP}$, excluding the unity coefficient. This prevents certain Pauli coefficients from being unintentionally maximized due to their absence from the cost function. The third term sums up the Pauli coefficients $w_{KK}$ of specific error channels that one aims to particularly suppress. These error channels have individual weightings $\gamma_K$, which can be chosen individually for a given task. 

This error-based gate optimization approach enables the targeted suppression of specific Pauli error channels. It is particularly well suited for designing gates with a noise bias or for gates involved in quantum error correction (QEC), where certain Pauli errors are more detrimental than others. We chose all Pauli channels that lead to fault tolerance braking errors to be in the specific set of the third term in Eq. \eqref{eqn:new_cost}. In the following section, we describe how the weightings $\gamma_K$ are determined to prioritize the suppression of errors that are most harmful to the QEC process.

\section{Application to the surface code} \label{sec:application_on_the_surface_code_qec_protocol}

We now focus on the integration of our CZZ gate in a surface-code error correction protocol. A key characteristic of QEC codes, including the surface code, is the code distance $d$, defined as the minimum number of physical qubits that need to be faulty in order to create a non-detectable logical error. This means that a QEC code with a distance $d$ can detect errors that affect $d-1$ physical qubits, and can correct errors on at least $t = \lfloor \frac{d-1}{2} \rfloor$ faulty physical qubits. In the following we investigate two variants of the surface code: The unrotated surface code, shown in Fig.~\ref{fig:surface_stabilizer_readout} a) for $d=3$, which requires $2d^2+2(d-1)^2 -1 $ physical qubits to encode one logical qubit (including auxiliary qubits for stabilizer readout) and the more qubit-efficient rotated surface code, shown in Fig.~\ref{fig:surface_stabilizer_readout} b), which requires only $2d^2-1$ physical qubits. Because of the smaller qubit count, experimental realizations mainly focus on the rotated surface code, as illustrated by a number of recent experimental demonstrations on superconducting hardware\,\cite{Andersen2020,Krinner2022,Acharya2023}. 

In Ref.\,\cite{old2025faulttolerant} some of us have shown that -- contrary to the folklore that multi-qubit gates are incompatible with strictly fault-tolerant circuit designs -- the unrotated surface code can retain its fault-tolerance when employing a syndrome readout protocol based on 3-qubit CZZ gates. Fault-tolerance for our purposes means that a circuit is \emph{distance preserving} in the following sense: If we implement the stabilizer measurements of a distance-$d$ QEC code using a circuit with a circuit-level noise model where every location can be faulty, all up to weight-$t$ combinations of faults ('order $p^{t}$ faults') can still be corrected. 
Even with 3-qubit depolarizing noise applied after the CZZ gates, unrotated surface code stabilizer readout circuits remain strictly fault-tolerant and achieve a higher threshold than the standard protocol using successive two-qubit CZ gates. Rotated surface codes, however, have their circuit distance halved when allowing for arbitrary $3$-qubit depolarizing noise errors on the support of the 3-qubit CZZ gates. 

Here we improve the performance of the rotated surface code, by optimizing the CZZ gate in a fault-tolerance adapted way to better suppress fault-tolerance breaking errors for both the rotated and unrotated surface code. To that end, we first characterize which faults are particularly harmful in the rotated surface-code implementation.

\subsection{Fault-tolerance breaking faults}\label{sec:ft_breaking_faults}

The following analysis examines which of the $4^3-1=63$ non-identity Pauli terms on the support of 3-qubit gates are particularly harmful. Although a logical error is a global result of the protocol, we can still identify which local errors on individual gates lead to uncorrectable errors more frequently than others. For concreteness, we restrict ourselves to $5$ rounds of distance-$5$ rotated surface code syndrome measurement circuits, but the approach is extendible to larger distances.
For this analysis, we initialize a graph $\mathcal{G} = (\mathcal{V},\mathcal{E})$ with the Pauli terms as nodes $V$, $|V| = 63$, and
simulate noisy circuits by explicitly placing all order $p^2$ faults. 
As the rotated surface code with $\mathrm{CZZ}$ gates is not fault-tolerant, this set of two-fault processes includes faults that lead to an error syndrome which will lead to a logical error. 
Whenever the syndrome of a combination of Pauli operators $(P,P')$ at any combination of locations leads to an error-guess resulting in a logical error, we draw an edge $E = (P,P') \to \mathcal{E}$ between the respective nodes and call the Paulis \textit{conspiring}. We then record the occurrence probability of the fault combination, $p_{(P,P')} = p_{(P)} p_{(P')} \prod p_{(I)}$, where $p_{(P)}$ is the probability of Pauli string $P$ in the corresponding channel, $p_{(I)}$ is the probability that no fault occurs, and the product runs over all non-faulty locations. For a circuit with $n_{\mathrm{l}}$ locations and a uniform single-parameter noise model, one gets $p_{(P,P')} = \left(\frac{p}{63}\right)^2 (1-p)^{n_{\mathrm{l}}-2}$.
If an edge already exists, we add the occurrence probability to the respective record. 
In general, also order $p^2$ combinations of a CZZ fault and e.g. a single-qubit error can lead to a logical error. While we include all other error mechanisms in the occurrence probability, motivated by higher fidelities of single-qubit operations in current experiments, we only consider how the faults on the three-qubit gates conspire to focus on the harmfulness of the gate. 

\begin{figure*}
        \centering
        \includegraphics[width=\linewidth]{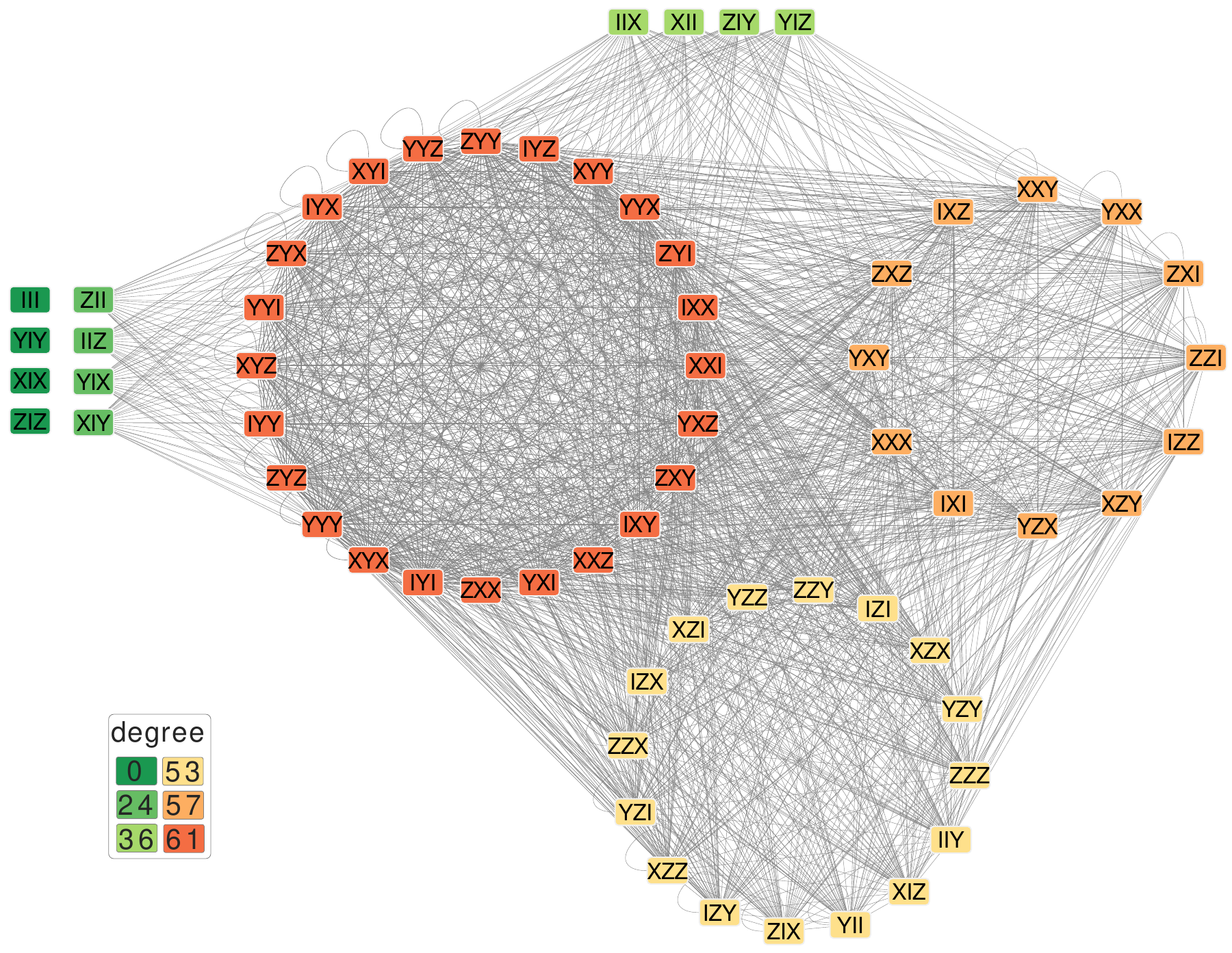}
        \caption{Graph $G$ of conspiring Pauli terms. We simulate $5$ rounds of distance-$5$ rotated surface code stabilizer measurement circuits in a $Z$-memory experiment (see main text). We manually place all possible order-$2$ combinations of faults of uniform $3$-qubit depolarizing channels after the $3$-qubit CZZ gates.
        Whenever a fault combination leads to a logical error, there is an edge in between the constituent Pauli terms. We divide the Pauli operators into groups with the same degree, indicating the harmfulness of specific Pauli terms. For example, four Pauli terms ($III, ZIZ, YIY, XIX)$ are never part of a failing combination, while others (e.g. $IYI$ or $XYX$) lead to a failure in any other case (i.e.~in $61$ cases including conspiring with itself).}
        \label{fig:paulis_graph}
\end{figure*}

We show the resulting graph for $p = 0.001$ in Fig.~\ref{fig:paulis_graph}. We find that few Pauli terms ($III, ZIZ, YIY, XIX)$ are never part of a conspiring combination, while others (e.g. $IYY$ or $XXI$) lead to a failure in any other (i.e. $61$) case.  
From the obtained data, we calculate single Pauli term marginals as the sum of failure probabilities at adjacent edges,
\begin{align}
    p_P = \sum_{P' : (P,P') \in \mathcal{E}} p_{(P,P')}. \label{eqn:Pauli_marginal}
\end{align}
We interpret these marginals as the average harmfulness of the respective Pauli term in a uniform $3$-qubit depolarizing channel on the gate. 
The marginals are displayed in Fig.~\ref{fig:failing_paulis_rotated_5_5_True_SI1000_2}. Notably, the harmfulness ranges over two orders of magnitude and correlates with the degree of the terms in the graph $G$. The largest contributions are for example $IYI, XYX, YYY$ and $ZYZ$. 

\subsection{Optimized Gates}

We now focus on the optimization of the logical QEC performance, therefore we normalize the Pauli marginals given by Eq.~\ref{eqn:Pauli_marginal} and use them as additional weights $\gamma_K$ for the Pauli channels $w_{KK}$ in the cost function defined in Eq.~\ref{eqn:new_cost}. These marginals correlate to the probabilities associated with each Pauli error channel contributing to a fault-tolerance-breaking error event. With this method we can specifically suppress harmful errors and boost the QEC performance. 
In the following will use the notation $\eta = [\eta_1,\eta_2,\eta_3]$ for the cost weightings.

\begin{figure*}
    \centering
    \includegraphics[width=\linewidth]{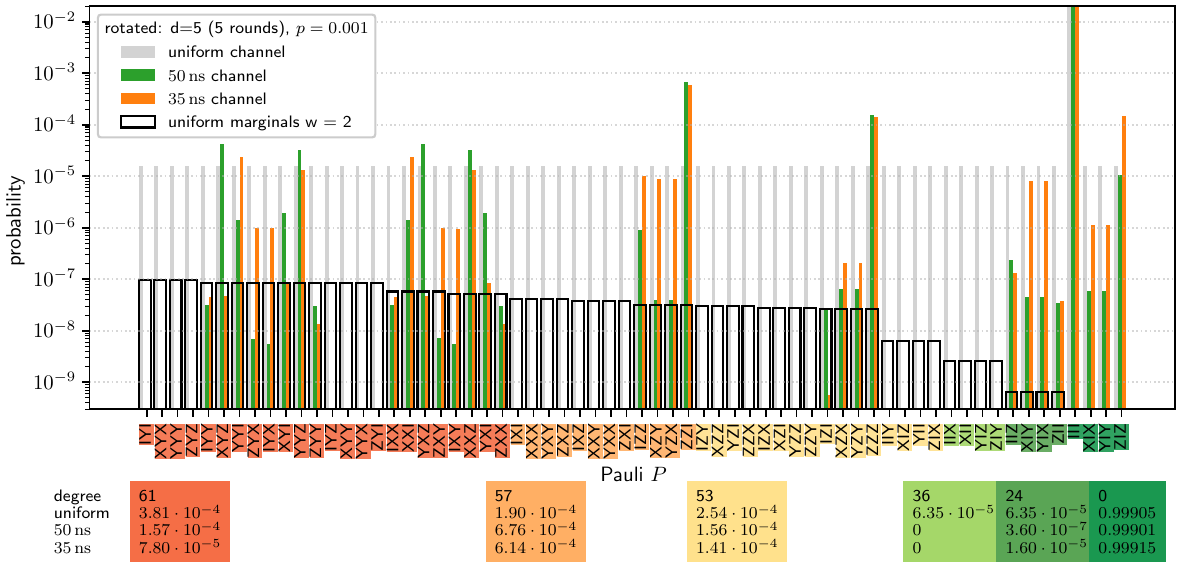}
    \caption{Effective Pauli channels are displayed by the orange bars for the $35\,$ns gate and the green bars for the $50\,$ns gate. The grey bars show the Pauli channels for a uniform noise distribution and the Pauli-term marginals, c.f. Eq. \eqref{eqn:Pauli_marginal} are depicted by the black outlined bars. All Pauli channels and Pauli marginals are rescaled to a physical error probability of $p=0.001$. 
    The Pauli-term marginals $p_P$ of order-$2$ faults are calculated from the occurrence probabilities, i.e.~we sum the outgoing edges of the nodes in the graph $G$ of Fig.\,\ref{fig:paulis_graph}, cf. Eqn.\,\ref{eqn:Pauli_marginal}. We sort the Pauli operators shown on the horizontal axis by decreasing marginal probability and color the background of their labels according to the degree of the corresponding node in $G$. Notably, the effective Pauli channels are structured in such a way that terms with an odd number of $X$- or $Y$- content are heavily suppressed. We also show the sums of the Pauli-channel contributions for terms with the same degree in $G$ at the bottom. Notably, Pauli terms with higher degrees are more strongly suppressed by the $35\,\mathrm{ns}$ gate than those with lower degrees.}
    \label{fig:failing_paulis_rotated_5_5_True_SI1000_2}
\end{figure*}

We will next discuss two possible pulses for the CZZ gate, a pulse with a duration of $35\,\mathrm{ns}$, obtained with the weightings $\eta = [1,10,0.1]$, and one with a duration of $50\,\mathrm{ns}$, obtained with  $\eta = [1,10,0.1]$. The respective gate and decoherence errors are listed in Table~\ref{tab:gate_errors}. For these gates, we calculate the effective Pauli channels as described in Sec.~\ref{subsec:error_model} and display them in Fig.~\ref{fig:failing_paulis_rotated_5_5_True_SI1000_2}. We can see that the Pauli terms with the largest uniform marginals (indicated by the bars on the left side of the plot, except for the "III"-entry) are suppressed compared to the other Pauli terms (bars in the right part of the plot).

\begin{table}[]
    \centering
    \renewcommand{\arraystretch}{1.5}
    \begin{tabular}{c|c|c}
        & $35\,$ns & $50\,$ns \\ \hline
        $\varepsilon_{\text{gate}}$ & $3.646\cdot10^{-4}$ & $6.529 \cdot 10^{-4}$\\ 
         $\varepsilon_{\text{dec}}$ & $6.9975\cdot 10^{-4}$ & $9.995 \cdot 10^{-4}$
    \end{tabular}
    \caption{Errors for the $35\,$ns and the $50\,$ns gate.}
    \label{tab:gate_errors}
\end{table}

\subsection{Improved surface code performance and resource analysis}
We conduct memory experiments in rotated and unrotated surface codes using \texttt{stim}\,\cite{Gidney2021}. For a distance-$d$ $Z$-($X$-)memory experiment, we prepare $\ket{0}^{\otimes n}$ ($\ket{+}^{\otimes n}$), measure the stabilizers $d$ times using the readout protocol shown in Fig.\,\ref{fig:surface_stabilizer_readout} and finally measure the data qubits in the $Z$ ($X$) basis. We simulate both bases and report the joint logical error rate as detailed in App.\,\ref{apx:memory_exp}.
The schedule for the rotated surface code is depicted in panel a), the protocol for the unrotated code can be seen in panel b). The readout for the $Z$-stabilizers is done in two steps, first all the CZZ gates connecting the south and west (SW) 
data qubits are executed. In the second step the north-east (NE) CZZ gates are performed. After the gates, the ancilla qubits are measured, while the CZZ gates used to map out the $X$-stabilizers are applied in a NW-SE ordering. For the readout of the $X$-stabilizers we additionally apply Hadamard gates before and after the CZZ gates on all qubits as depicted in Fig.\,\ref{fig:surface_stabilizer_readout} c).

We use a circuit level noise model inspired by the SI1000 model of Ref.\,\cite{Gidney2021}, where each circuit location can be faulty. Single-qubit initializations (measurements) in the $Z$-basis are followed (preceded) by a bit-flip channel of strengths $p_{\mathrm{reset}} = 2p$, $p_{\mathrm{measure}} = 5p$. Single- and two- qubit gates are modeled as ideal followed by a single- and two- qubit depolarizing channels of strength $p$. For three-qubit gates we compare gates followed by three-qubit uniform depolarizing channels against gates followed by the noise channel derived for our gate, see also Sec.\,\ref{sec:ft_breaking_faults}. Additionally, we assume idling noise for each waiting location of $p_{\mathrm{idle}} = 0.1 p$.

\begin{figure*}
    \centering
    \includegraphics[width=\linewidth]{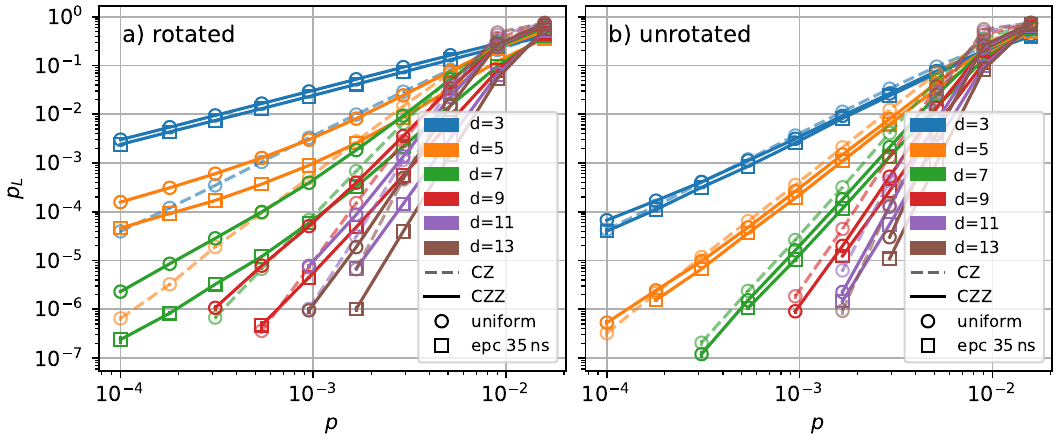}
    \caption{Results for rotated (a) and unrotated (b) surface codes: We show the logical error rates of uniform depolarizing error models (circles) and of the effective Pauli channel (epc) of the $35\,\mathrm{ns}-[1,10,0.1]$ gate (squares) for different surface code distances displayed in the legend. We include an idling noise strength of $0.1p$. With dashed lines, we show the logical error rates using a fault-tolerant readout with 2-qubit CZ gates. a) The CZZ protocol for the rotated surface code is represented by the continuous lines. It is not fault tolerant, such that for small error rates, the curves show a reduced scaling. However, for larger physical error rates the CZZ protocol has a lower logical error rate, even with a uniform depolarizing noise model. With the effective error model, this regime extends to smaller physical error rates, such that for distance $11$, e.g., only for error rates $<10^{-3}$, the CZ protocol outperforms the CZZ protocol. b) For unrotated surface codes, the CZZ protocol is fault tolerant and therefore outperforms the CZ protocol for almost all simulated physical error rates. The effective error model only marginally improves upon the uniform depolarizing model.}
    \label{fig:plot_rot_unrot_cz_czz_epc_idle_log}
\end{figure*}

The results are shown in Fig.\,\ref{fig:plot_rot_unrot_cz_czz_epc_idle_log}, where we compare the error correction performance of the surface codes implementing the CZZ gate readout protocols. 
We compare the two noise models obtained by optimal control with the uniform noise model. For the unrotated surface codes depicted in Fig.\, \ref{fig:plot_rot_unrot_cz_czz_epc_idle_log} b), we observe a small improvement in logical error rate. Due to the fault-tolerance of the circuits, the scaling of the logical error rate in the low error regime is in any case $\propto p^{t+1}$. Small deviations from the expected scaling are related to sub-optimal decomposition of faults into matchable faults, see also Ref.\,\cite{old2025faulttolerant}. This suggests that optimizing pulses to suppress harmful errors can also enhance standard error correction readout schemes that use two-qubit gates.

\begin{figure*}
    \centering
    \includegraphics[width=\linewidth]{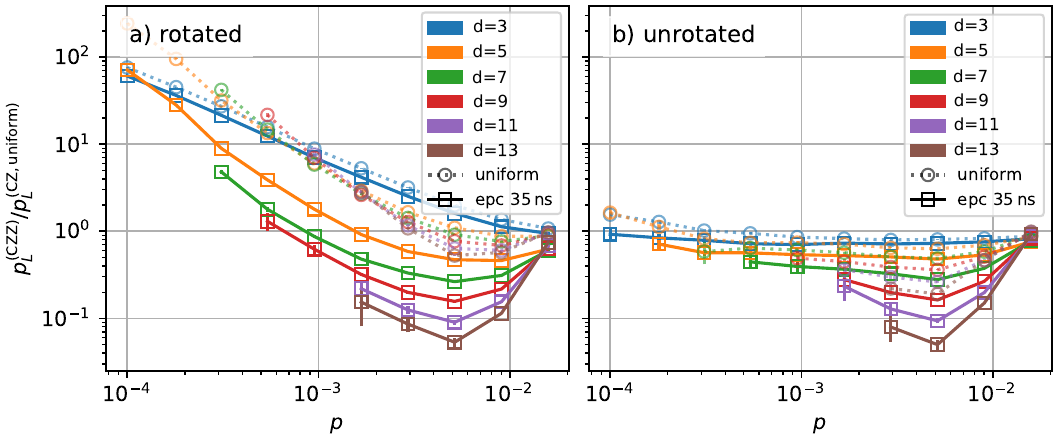}
    \caption{Logical error rates for CZZ circuits for the uniform gate noise model and the $35\,\mathrm{ns}-[1,10,0.1]$ gate with corresponding effective Pauli error model (epc) depicted by the squares and the continuous lines compared to the uniform depolarizing channel on $\mathrm{CZ}$ gates represented by the dashed lines with circles. The numbers in the legend display the respective surface code distances. When $p_L^{(\mathrm{(CZZ)}}/p_L^{(\mathrm{(CZ, uniform)}} < 1$, there is an advantage in using the $\mathrm{CZZ}$ gate readout. a) For the rotated surface codes and a uniform depolarizing noise CZZ protocol, there is an advantage for larger physical error rates. The regime of advantage shifts to smaller logical error rates when using the effective error model. b) For the unrotated surface codes, both uniform and effective error model on the CZZ protocol outperform the uniform CZ protocol for most simulated physical error rates. The effective error model only marginally improves upon the uniform depolarizing model.}
    \label{fig:plot_rot_unrot_cz_by_czz_epc_idle_log}
\end{figure*}

For the rotated surface code, displayed in Fig.\,\ref{fig:plot_rot_unrot_cz_czz_epc_idle_log} a), the effect of the suppression of fault-tolerance breaking errors is much more distinct. Here the improvement amounts to about one order of magnitude for higher code distances. Since the rotated surface code doesn't have a fault-tolerant readout schedule for the CZZ gate, the logical error scaling flattens out with a smaller physical error rate. We observe a scaling $p_L \propto p^{\lceil \frac{d-1}{2}}\rceil$. The expected scaling from the non-FT distance $\frac{d+1}{2}$ is $\lfloor \frac{d-1}{2} \rfloor + 1$ which can be observed around $p \approx 10^{-3}$. We attribute the further reduction for very low physical error rates again to sub-optimal fault decompositions. Nevertheless, in the experimentally relevant regime of $p \in [10^{-3}\dots10^{-2}]$\,\cite{Acharya2024}, the improvement of the error-suppressed CZZ gates is significant. This becomes even more evident if we look at the comparison of the gain in logical error rate, which is displayed in Fig.\,\ref{fig:plot_rot_unrot_cz_by_czz_epc_idle_log}.

\begin{figure*}
    \centering
    \includegraphics[width=\linewidth]{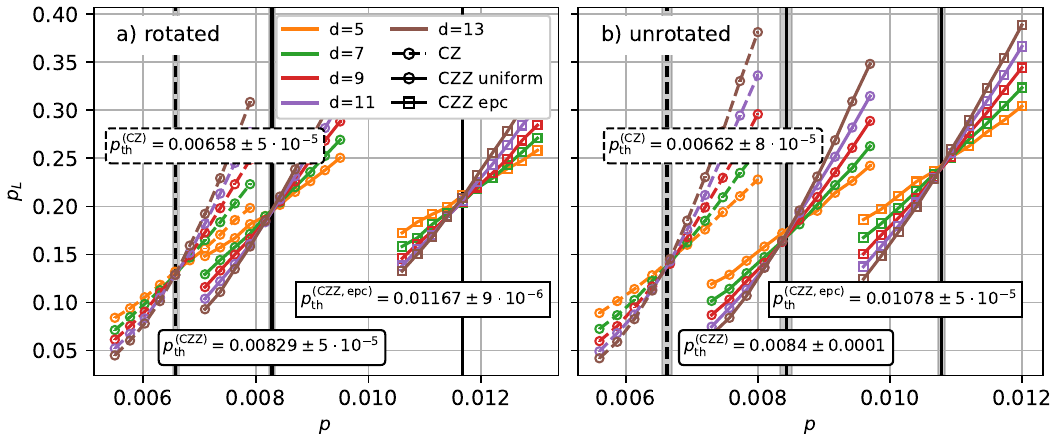}
    \caption{Threshold plots. a) Rotated surface codes show an increase in threshold from $p_{\mathrm{th}}^{(\mathrm{CZ})} \approx 0.66\%$ to $p_{\mathrm{th}}^{(\mathrm{CZZ})} \approx 0.83\%$ and $p_{\mathrm{th}}^{(\mathrm{CZZ, epc})} \approx 1.2\%$. b) Unrotated surface codes show an increase in threshold from $p_{\mathrm{th}}^{(\mathrm{CZ})} \approx 0.66\%$ to $p_{\mathrm{th}}^{(\mathrm{CZZ})} \approx 0.84\%$ and $p_{\mathrm{th}}^{(\mathrm{CZZ, epc})} \approx 1.1\%$. The colors encode different code distances, circles refer to uniform noise and squares to effective Pauli channel (epc) noise, see legend. Dashed lines mark circuits with two-qubit gates, while solid lines mark circuits using the CZZ gate. Uncertainties are obtained from a finite size scaling analysis, detailed in the appendix.}
    \label{fig:threshold_plots}
\end{figure*}

Our simulations also show that the new CZZ gate readout schedules improve the physical qubit error threshold from $\approx 0.66\,\%$ to $\approx 1.2\,\%$ for the rotated surface code, and from $\approx 0.66\,\%$ to $ \approx 1.1\,\%$ for the unrotated surface code, as can be seen in Fig.\,\ref{fig:threshold_plots}. The significant increase in physical error threshold can be attributed to the reduced number of fault locations, as explained in detail in Ref.\,\cite{old2025faulttolerant}. Note that these results can be regarded as an upper bound for the physical threshold and logical error reduction, since our error model does not include all errors present in an experimental realization.

\begin{figure*}
    \centering
    \includegraphics[width=\linewidth]{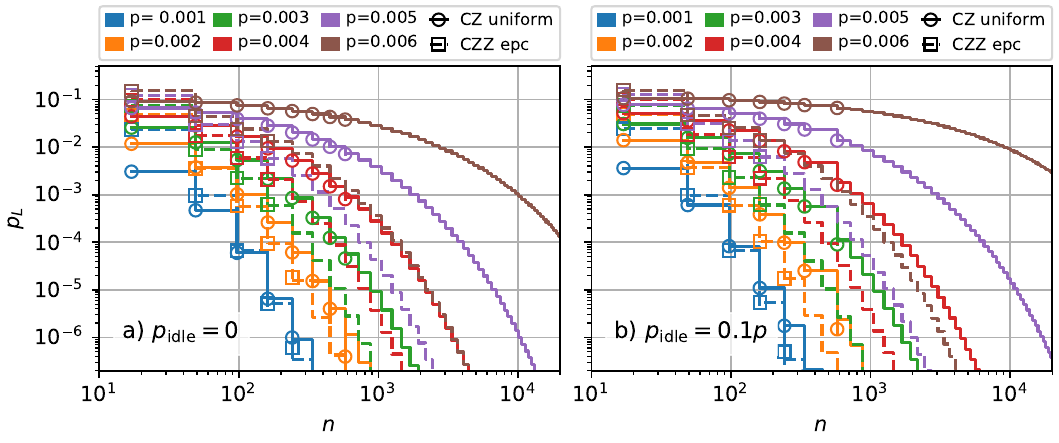}
    \caption{Number of physical qubits required to reach a target logical error rate, including auxiliary qubits, i.e. $n = 2d^2-1$. We compare rotated surface code implemented with CZ gates and a uniform depolarizing error model (continuous lines and circles), to rotated surface code implemented with CZZ gates and the effective Pauli channel of the $35\,\mathrm{ns}$ gate (dashed lines and squares). We plot the resource estimations for different physical error probabilities from $p=0.1\%$ to $p=0.6\%$, see legend for color coding. a) No idling noise. For example, to reach a logical error rate of $p_L = 10^{-6}$ at $p = 0.3\%$, $1457$ physical qubits (distance $27$) are required for the CZ protocol, but only $1057$ physical qubits (distance $23$) with the CZZ protocol, an advantage of $\approx 27.5 \%$. 
    b) With idling noise of strength $p_{\mathrm{idle}} = 0.1 p$ this reduction increases. The same $1057$ physical qubits (distance $23$) CZZ protocol uses $\approx 37.1\%$ less qubits than the CZ protocols with $1681$ qubits (distance $29$).}
    \label{fig:plot_p_L_n_p_noise_model.pdf}
\end{figure*}

We now compare the resource requirements for a fault-tolerant CZ readout protocol under uniform depolarizing noise and the optimized CZZ gate with the effective Pauli noise model as the underlying noise channel on 3-qubit gates. 
For that, we calculate the number of physical qubits required to reach a target logical error rate. We plot the logical error rate for different physical errors $p$ over the number of physical qubits (including ancilla qubits) in Fig.\,\ref{fig:plot_p_L_n_p_noise_model.pdf}. We investigate these graphs for no noise on idling qubits in Fig.\,\ref{fig:plot_p_L_n_p_noise_model.pdf} a) (left) and for a noise of $p_{\mathrm{idle}} = 0.1 p$ on idling qubits in Fig.\,\ref{fig:plot_p_L_n_p_noise_model.pdf} b) (right). 
We fit the function
\begin{align}
    p_L(n) = c_0 \left(\frac{p}{c_1}\right)^{c_2 \sqrt{n}}
\end{align}
to the simulated data to allow for an extrapolation to large surface codes and plot a step function that rounds to the next larger surface code available. 
In both simulations and and in particular for physical error rates $\geq 0.3\%$, the CZZ-gate based readout protocols requires fewer physical qubits to reach a target logical error rate. 
For example, reaching a logical error rate of $p_L = 10^{-6}$ at $p = 0.3\%$ requires $1457$ physical qubits (distance $27$) for the CZ protocol, but only $1057$ physical qubits (distance $23$) with the CZZ protocol, an advantage of $\approx 27.5 \%$. 
This reduction increases if we include idling noise, where the same $1057$ physical qubits (distance $23$) CZZ protocol uses $\approx 37.1\%$ less qubits than the CZ protocols with $1681$ (distance $29$) qubits.  

\section{Conclusion and outlook}
In this paper we present a new surface-code stabilizer readout protocol for the rotated and unrotated surface-code using CZZ gates. We design our CZZ gate by using optimal control, to strongly suppress Pauli errors that are explicitly harmful for the surface-code QEC. We show that although the CZZ gate readout protocol is not fault tolerant for the rotated surface-code, in the experimental achievable regime, it outperforms the common two-qubit readout protocol with two-qubit gates in physical error threshold and logical error rate scaling. For the unrotated surface code some of us have shown in a parallel work\,\cite{old2025faulttolerant} that a fault-tolerant readout schedule using CZZ gates can be found. We show that even for the fault-tolerant readout schedule our new readout protocol improves the unrotated surface-code QEC slightly in logical error scaling and significantly in the physical qubit threshold.

These results suggest that shifting the gate optimization to suppress fault-tolerance-breaking Pauli errors could also improve conventional fault-tolerant two-qubit gate readout procedures, such as those used by, e.g., Google\,\cite{Acharya2023,Acharya2024}, ETH Zürich\,\cite{Krinner2022} and others. Beyond targeting specific uncorrectable quantum error correction (QEC) errors, the gate design approach described here can also be applied to the design of other quantum gates, also beyond superconducting qubit platforms, and particularly gates with a strong noise bias, since it enables stronger suppression of individual Pauli error channels.

Based on the  structure of the underlying superconducting circuit, consisting of only fixed-frequency and tunable transmon qubits, the proposed stabilizer readout scheme using CZZ gates is well-suited for an experimental implementation on existing superconducting circuit hardware, without changing the hardware design. This makes the performance of the CZZ gate/ readout protocol on real hardware an interesting research question. Another promising research direction emerging from this work is the implementation of targeted Pauli channel suppression in quantum gate design.

\section{Acknowledgements}
This research is part of the Munich Quantum Valley (K-8), which is supported by the Bavarian state government with funds from the Hightech Agenda Bayern Plus. We additionally acknowledge support by the BMBF projects GeQCoS (Grant No. 13N15684) and MUNIQC-ATOMS (Grant No. 13N16070). Furthermore, J.O. and M.M. acknowledge support by the
European Union’s Horizon Europe research and innovation programme under Grant Agreement No. 101114305
(“MILLENION-SGA1” EU Project) and by the Deutsche
Forschungsgemeinschaft (DFG, German Research Foundation) under Germany’s Excellence Strategy “Cluster
of Excellence Matter and Light for Quantum Computing
(ML4Q) EXC 2004/1” 390534769 and the ERC Start-
ing Grant QNets through Grant No. 804247. 
The authors gratefully acknowledge the computing time provided to them at the NHR Center NHR4CES at RWTH Aachen University (Project No. p0020074). This is funded by the Federal Ministry of Education and Research and the state governments participating on the basis of the resolutions of the GWK for national high performance computing at universities.
\FloatBarrier

\bibliography{sources.bib}

\begin{thebibliography}{47}%
\makeatletter
\providecommand \@ifxundefined [1]{%
 \@ifx{#1\undefined}
}%
\providecommand \@ifnum [1]{%
 \ifnum #1\expandafter \@firstoftwo
 \else \expandafter \@secondoftwo
 \fi
}%
\providecommand \@ifx [1]{%
 \ifx #1\expandafter \@firstoftwo
 \else \expandafter \@secondoftwo
 \fi
}%
\providecommand \natexlab [1]{#1}%
\providecommand \enquote  [1]{``#1''}%
\providecommand \bibnamefont  [1]{#1}%
\providecommand \bibfnamefont [1]{#1}%
\providecommand \citenamefont [1]{#1}%
\providecommand \href@noop [0]{\@secondoftwo}%
\providecommand \href [0]{\begingroup \@sanitize@url \@href}%
\providecommand \@href[1]{\@@startlink{#1}\@@href}%
\providecommand \@@href[1]{\endgroup#1\@@endlink}%
\providecommand \@sanitize@url [0]{\catcode `\\12\catcode `\$12\catcode `\&12\catcode `\#12\catcode `\^12\catcode `\_12\catcode `\%12\relax}%
\providecommand \@@startlink[1]{}%
\providecommand \@@endlink[0]{}%
\providecommand \url  [0]{\begingroup\@sanitize@url \@url }%
\providecommand \@url [1]{\endgroup\@href {#1}{\urlprefix }}%
\providecommand \urlprefix  [0]{URL }%
\providecommand \Eprint [0]{\href }%
\providecommand \doibase [0]{http://dx.doi.org/}%
\providecommand \selectlanguage [0]{\@gobble}%
\providecommand \bibinfo  [0]{\@secondoftwo}%
\providecommand \bibfield  [0]{\@secondoftwo}%
\providecommand \translation [1]{[#1]}%
\providecommand \BibitemOpen [0]{}%
\providecommand \bibitemStop [0]{}%
\providecommand \bibitemNoStop [0]{.\EOS\space}%
\providecommand \EOS [0]{\spacefactor3000\relax}%
\providecommand \BibitemShut  [1]{\csname bibitem#1\endcsname}%
\let\auto@bib@innerbib\@empty
\bibitem [{\citenamefont {Dalzell}\ \emph {et~al.}(2025)\citenamefont {Dalzell}, \citenamefont {McArdle}, \citenamefont {Berta}, \citenamefont {Bienias},\ and\ \citenamefont {Chen}}]{Dalzell2025}%
  \BibitemOpen
  \bibfield  {author} {\bibinfo {author} {\bibfnamefont {A.~M.}\ \bibnamefont {Dalzell}}, \bibinfo {author} {\bibfnamefont {S.}~\bibnamefont {McArdle}}, \bibinfo {author} {\bibfnamefont {M.}~\bibnamefont {Berta}}, \bibinfo {author} {\bibfnamefont {P.}~\bibnamefont {Bienias}}, \ and\ \bibinfo {author} {\bibfnamefont {C.-F.}\ \bibnamefont {Chen}},\ }\href@noop {} {\emph {\bibinfo {title} {Quantum Algorithms A Survey of Applications and End-To-end Complexities}}}\ (\bibinfo  {publisher} {Cambridge University Press},\ \bibinfo {year} {2025})\BibitemShut {NoStop}%
\bibitem [{\citenamefont {Kitaev}(2003)}]{Kitaev2003}%
  \BibitemOpen
  \bibfield  {author} {\bibinfo {author} {\bibfnamefont {A.}~\bibnamefont {Kitaev}},\ }\href {\doibase 10.1016/s0003-4916(02)00018-0} {\bibfield  {journal} {\bibinfo  {journal} {Annals of Physics}\ }\textbf {\bibinfo {volume} {303}},\ \bibinfo {pages} {2} (\bibinfo {year} {2003})}\BibitemShut {NoStop}%
\bibitem [{\citenamefont {Kitaev}(1997)}]{Kitaev1997}%
  \BibitemOpen
  \bibfield  {author} {\bibinfo {author} {\bibfnamefont {A.~Y.}\ \bibnamefont {Kitaev}},\ }\enquote {\bibinfo {title} {Quantum error correction with imperfect gates},}\ in\ \href {\doibase 10.1007/978-1-4615-5923-8_19} {\emph {\bibinfo {booktitle} {Quantum Communication, Computing, and Measurement}}}\ (\bibinfo  {publisher} {Springer US},\ \bibinfo {year} {1997})\ pp.\ \bibinfo {pages} {181--188}\BibitemShut {NoStop}%
\bibitem [{\citenamefont {Bravyi}\ and\ \citenamefont {Kitaev}(1998)}]{Bravyi1998}%
  \BibitemOpen
  \bibfield  {author} {\bibinfo {author} {\bibfnamefont {S.~B.}\ \bibnamefont {Bravyi}}\ and\ \bibinfo {author} {\bibfnamefont {A.~Y.}\ \bibnamefont {Kitaev}},\ }\href {\doibase https://doi.org/10.48550/arXiv.quant-ph/9811052} {\  (\bibinfo {year} {1998}),\ https://doi.org/10.48550/arXiv.quant-ph/9811052},\ \Eprint {http://arxiv.org/abs/quant-ph/9811052} {arXiv:quant-ph/9811052 [quant-ph]} \BibitemShut {NoStop}%
\bibitem [{\citenamefont {Dennis}\ \emph {et~al.}(2002)\citenamefont {Dennis}, \citenamefont {Kitaev}, \citenamefont {Landahl},\ and\ \citenamefont {Preskill}}]{Dennis2002}%
  \BibitemOpen
  \bibfield  {author} {\bibinfo {author} {\bibfnamefont {E.}~\bibnamefont {Dennis}}, \bibinfo {author} {\bibfnamefont {A.}~\bibnamefont {Kitaev}}, \bibinfo {author} {\bibfnamefont {A.}~\bibnamefont {Landahl}}, \ and\ \bibinfo {author} {\bibfnamefont {J.}~\bibnamefont {Preskill}},\ }\href {\doibase 10.1063/1.1499754} {\bibfield  {journal} {\bibinfo  {journal} {Journal of Mathematical Physics}\ }\textbf {\bibinfo {volume} {43}},\ \bibinfo {pages} {4452} (\bibinfo {year} {2002})}\BibitemShut {NoStop}%
\bibitem [{\citenamefont {Fowler}\ \emph {et~al.}(2012)\citenamefont {Fowler}, \citenamefont {Mariantoni}, \citenamefont {Martinis},\ and\ \citenamefont {Cleland}}]{Fowler2012}%
  \BibitemOpen
  \bibfield  {author} {\bibinfo {author} {\bibfnamefont {A.~G.}\ \bibnamefont {Fowler}}, \bibinfo {author} {\bibfnamefont {M.}~\bibnamefont {Mariantoni}}, \bibinfo {author} {\bibfnamefont {J.~M.}\ \bibnamefont {Martinis}}, \ and\ \bibinfo {author} {\bibfnamefont {A.~N.}\ \bibnamefont {Cleland}},\ }\href {\doibase 10.1103/physreva.86.032324} {\bibfield  {journal} {\bibinfo  {journal} {Physical Review A}\ }\textbf {\bibinfo {volume} {86}},\ \bibinfo {pages} {032324} (\bibinfo {year} {2012})}\BibitemShut {NoStop}%
\bibitem [{\citenamefont {Wang}\ \emph {et~al.}(2011)\citenamefont {Wang}, \citenamefont {Fowler},\ and\ \citenamefont {Hollenberg}}]{Wang2011}%
  \BibitemOpen
  \bibfield  {author} {\bibinfo {author} {\bibfnamefont {D.~S.}\ \bibnamefont {Wang}}, \bibinfo {author} {\bibfnamefont {A.~G.}\ \bibnamefont {Fowler}}, \ and\ \bibinfo {author} {\bibfnamefont {L.~C.~L.}\ \bibnamefont {Hollenberg}},\ }\href {\doibase 10.1103/physreva.83.020302} {\bibfield  {journal} {\bibinfo  {journal} {Physical Review A}\ }\textbf {\bibinfo {volume} {83}},\ \bibinfo {pages} {020302} (\bibinfo {year} {2011})}\BibitemShut {NoStop}%
\bibitem [{\citenamefont {Andersen}\ \emph {et~al.}(2020)\citenamefont {Andersen}, \citenamefont {Remm}, \citenamefont {Lazar}, \citenamefont {Krinner}, \citenamefont {Lacroix}, \citenamefont {Norris}, \citenamefont {Gabureac}, \citenamefont {Eichler},\ and\ \citenamefont {Wallraff}}]{Andersen2020}%
  \BibitemOpen
  \bibfield  {author} {\bibinfo {author} {\bibfnamefont {C.~K.}\ \bibnamefont {Andersen}}, \bibinfo {author} {\bibfnamefont {A.}~\bibnamefont {Remm}}, \bibinfo {author} {\bibfnamefont {S.}~\bibnamefont {Lazar}}, \bibinfo {author} {\bibfnamefont {S.}~\bibnamefont {Krinner}}, \bibinfo {author} {\bibfnamefont {N.}~\bibnamefont {Lacroix}}, \bibinfo {author} {\bibfnamefont {G.~J.}\ \bibnamefont {Norris}}, \bibinfo {author} {\bibfnamefont {M.}~\bibnamefont {Gabureac}}, \bibinfo {author} {\bibfnamefont {C.}~\bibnamefont {Eichler}}, \ and\ \bibinfo {author} {\bibfnamefont {A.}~\bibnamefont {Wallraff}},\ }\href {\doibase 10.1038/s41567-020-0920-y} {\bibfield  {journal} {\bibinfo  {journal} {Nature Physics}\ }\textbf {\bibinfo {volume} {16}},\ \bibinfo {pages} {875} (\bibinfo {year} {2020})}\BibitemShut {NoStop}%
\bibitem [{\citenamefont {Krinner}\ \emph {et~al.}(2022)\citenamefont {Krinner}, \citenamefont {Lacroix}, \citenamefont {Remm}, \citenamefont {Di~Paolo}, \citenamefont {Genois}, \citenamefont {Leroux}, \citenamefont {Hellings}, \citenamefont {Lazar}, \citenamefont {Swiadek}, \citenamefont {Herrmann}, \citenamefont {Norris}, \citenamefont {Andersen}, \citenamefont {Müller}, \citenamefont {Blais}, \citenamefont {Eichler},\ and\ \citenamefont {Wallraff}}]{Krinner2022}%
  \BibitemOpen
  \bibfield  {author} {\bibinfo {author} {\bibfnamefont {S.}~\bibnamefont {Krinner}}, \bibinfo {author} {\bibfnamefont {N.}~\bibnamefont {Lacroix}}, \bibinfo {author} {\bibfnamefont {A.}~\bibnamefont {Remm}}, \bibinfo {author} {\bibfnamefont {A.}~\bibnamefont {Di~Paolo}}, \bibinfo {author} {\bibfnamefont {E.}~\bibnamefont {Genois}}, \bibinfo {author} {\bibfnamefont {C.}~\bibnamefont {Leroux}}, \bibinfo {author} {\bibfnamefont {C.}~\bibnamefont {Hellings}}, \bibinfo {author} {\bibfnamefont {S.}~\bibnamefont {Lazar}}, \bibinfo {author} {\bibfnamefont {F.}~\bibnamefont {Swiadek}}, \bibinfo {author} {\bibfnamefont {J.}~\bibnamefont {Herrmann}}, \bibinfo {author} {\bibfnamefont {G.~J.}\ \bibnamefont {Norris}}, \bibinfo {author} {\bibfnamefont {C.~K.}\ \bibnamefont {Andersen}}, \bibinfo {author} {\bibfnamefont {M.}~\bibnamefont {Müller}}, \bibinfo {author} {\bibfnamefont {A.}~\bibnamefont {Blais}}, \bibinfo {author} {\bibfnamefont {C.}~\bibnamefont {Eichler}}, \ and\ \bibinfo {author} {\bibfnamefont
  {A.}~\bibnamefont {Wallraff}},\ }\href {\doibase 10.1038/s41586-022-04566-8} {\bibfield  {journal} {\bibinfo  {journal} {Nature}\ }\textbf {\bibinfo {volume} {605}},\ \bibinfo {pages} {669} (\bibinfo {year} {2022})}\BibitemShut {NoStop}%
\bibitem [{\citenamefont {AI}(2023)}]{Acharya2023}%
  \BibitemOpen
  \bibfield  {author} {\bibinfo {author} {\bibfnamefont {G.~Q.}\ \bibnamefont {AI}},\ }\href {\doibase 10.1038/s41586-022-05434-1} {\bibfield  {journal} {\bibinfo  {journal} {Nature}\ }\textbf {\bibinfo {volume} {614}},\ \bibinfo {pages} {676} (\bibinfo {year} {2023})}\BibitemShut {NoStop}%
\bibitem [{\citenamefont {Acharya}(2024)}]{Acharya2024}%
  \BibitemOpen
  \bibfield  {author} {\bibinfo {author} {\bibfnamefont {R.~e.~a.}\ \bibnamefont {Acharya}},\ }\href {\doibase 10.1038/s41586-024-08449-y} {\bibfield  {journal} {\bibinfo  {journal} {Nature}\ }\textbf {\bibinfo {volume} {638}},\ \bibinfo {pages} {920} (\bibinfo {year} {2024})}\BibitemShut {NoStop}%
\bibitem [{\citenamefont {Besedin}\ \emph {et~al.}(2025)\citenamefont {Besedin}, \citenamefont {Kerschbaum}, \citenamefont {Knoll}, \citenamefont {Hesner}, \citenamefont {Bödeker}, \citenamefont {Colmenarez}, \citenamefont {Hofele}, \citenamefont {Lacroix}, \citenamefont {Hellings}, \citenamefont {Swiadek}, \citenamefont {Flasby}, \citenamefont {Panah}, \citenamefont {Zanuz}, \citenamefont {Müller},\ and\ \citenamefont {Wallraff}}]{Besedin2025}%
  \BibitemOpen
  \bibfield  {author} {\bibinfo {author} {\bibfnamefont {I.}~\bibnamefont {Besedin}}, \bibinfo {author} {\bibfnamefont {M.}~\bibnamefont {Kerschbaum}}, \bibinfo {author} {\bibfnamefont {J.}~\bibnamefont {Knoll}}, \bibinfo {author} {\bibfnamefont {I.}~\bibnamefont {Hesner}}, \bibinfo {author} {\bibfnamefont {L.}~\bibnamefont {Bödeker}}, \bibinfo {author} {\bibfnamefont {L.}~\bibnamefont {Colmenarez}}, \bibinfo {author} {\bibfnamefont {L.}~\bibnamefont {Hofele}}, \bibinfo {author} {\bibfnamefont {N.}~\bibnamefont {Lacroix}}, \bibinfo {author} {\bibfnamefont {C.}~\bibnamefont {Hellings}}, \bibinfo {author} {\bibfnamefont {F.}~\bibnamefont {Swiadek}}, \bibinfo {author} {\bibfnamefont {A.}~\bibnamefont {Flasby}}, \bibinfo {author} {\bibfnamefont {M.~B.}\ \bibnamefont {Panah}}, \bibinfo {author} {\bibfnamefont {D.~C.}\ \bibnamefont {Zanuz}}, \bibinfo {author} {\bibfnamefont {M.}~\bibnamefont {Müller}}, \ and\ \bibinfo {author} {\bibfnamefont {A.}~\bibnamefont {Wallraff}},\ }\href {\doibase
  https://doi.org/10.48550/arXiv.2501.04612} {\enquote {\bibinfo {title} {Realizing lattice surgery on two distance-three repetition codes with superconducting qubits},}\ } (\bibinfo {year} {2025})\BibitemShut {NoStop}%
\bibitem [{\citenamefont {DiVincenzo}\ and\ \citenamefont {Solgun}(2013)}]{DiVincenzo2013}%
  \BibitemOpen
  \bibfield  {author} {\bibinfo {author} {\bibfnamefont {D.~P.}\ \bibnamefont {DiVincenzo}}\ and\ \bibinfo {author} {\bibfnamefont {F.}~\bibnamefont {Solgun}},\ }\href {\doibase 10.1088/1367-2630/15/7/075001} {\bibfield  {journal} {\bibinfo  {journal} {New Journal of Physics}\ }\textbf {\bibinfo {volume} {15}},\ \bibinfo {pages} {075001} (\bibinfo {year} {2013})}\BibitemShut {NoStop}%
\bibitem [{\citenamefont {Ciani}\ and\ \citenamefont {DiVincenzo}(2017)}]{Ciani2017}%
  \BibitemOpen
  \bibfield  {author} {\bibinfo {author} {\bibfnamefont {A.}~\bibnamefont {Ciani}}\ and\ \bibinfo {author} {\bibfnamefont {D.~P.}\ \bibnamefont {DiVincenzo}},\ }\href {\doibase 10.1103/physrevb.96.214511} {\bibfield  {journal} {\bibinfo  {journal} {Physical Review B}\ }\textbf {\bibinfo {volume} {96}},\ \bibinfo {pages} {214511} (\bibinfo {year} {2017})}\BibitemShut {NoStop}%
\bibitem [{\citenamefont {Schwerdt}\ \emph {et~al.}(2022)\citenamefont {Schwerdt}, \citenamefont {Shapira}, \citenamefont {Manovitz},\ and\ \citenamefont {Ozeri}}]{Schwerdt2022}%
  \BibitemOpen
  \bibfield  {author} {\bibinfo {author} {\bibfnamefont {D.}~\bibnamefont {Schwerdt}}, \bibinfo {author} {\bibfnamefont {Y.}~\bibnamefont {Shapira}}, \bibinfo {author} {\bibfnamefont {T.}~\bibnamefont {Manovitz}}, \ and\ \bibinfo {author} {\bibfnamefont {R.}~\bibnamefont {Ozeri}},\ }\href {\doibase 10.1103/physreva.105.022612} {\bibfield  {journal} {\bibinfo  {journal} {Physical Review A}\ }\textbf {\bibinfo {volume} {105}},\ \bibinfo {pages} {022612} (\bibinfo {year} {2022})}\BibitemShut {NoStop}%
\bibitem [{\citenamefont {Üstün}\ \emph {et~al.}(2024)\citenamefont {Üstün}, \citenamefont {Morello},\ and\ \citenamefont {Devitt}}]{Uestuen2024}%
  \BibitemOpen
  \bibfield  {author} {\bibinfo {author} {\bibfnamefont {G.}~\bibnamefont {Üstün}}, \bibinfo {author} {\bibfnamefont {A.}~\bibnamefont {Morello}}, \ and\ \bibinfo {author} {\bibfnamefont {S.}~\bibnamefont {Devitt}},\ }\href {\doibase 10.1088/2058-9565/ad473c} {\bibfield  {journal} {\bibinfo  {journal} {Quantum Science and Technology}\ }\textbf {\bibinfo {volume} {9}},\ \bibinfo {pages} {035037} (\bibinfo {year} {2024})}\BibitemShut {NoStop}%
\bibitem [{\citenamefont {Reagor}\ \emph {et~al.}(2022)\citenamefont {Reagor}, \citenamefont {Bohdanowicz}, \citenamefont {Perez}, \citenamefont {Sete},\ and\ \citenamefont {Zeng}}]{Reagor2022}%
  \BibitemOpen
  \bibfield  {author} {\bibinfo {author} {\bibfnamefont {M.~J.}\ \bibnamefont {Reagor}}, \bibinfo {author} {\bibfnamefont {T.~C.}\ \bibnamefont {Bohdanowicz}}, \bibinfo {author} {\bibfnamefont {D.~R.}\ \bibnamefont {Perez}}, \bibinfo {author} {\bibfnamefont {E.~A.}\ \bibnamefont {Sete}}, \ and\ \bibinfo {author} {\bibfnamefont {W.~J.}\ \bibnamefont {Zeng}},\ }\href {\doibase 10.48550/ARXIV.2211.06382} {\enquote {\bibinfo {title} {Hardware optimized parity check gates for superconducting surface codes},}\ } (\bibinfo {year} {2022})\BibitemShut {NoStop}%
\bibitem [{\citenamefont {Cerfontaine}\ \emph {et~al.}(2020)\citenamefont {Cerfontaine}, \citenamefont {Otten},\ and\ \citenamefont {Bluhm}}]{Cerfontaine2020}%
  \BibitemOpen
  \bibfield  {author} {\bibinfo {author} {\bibfnamefont {P.}~\bibnamefont {Cerfontaine}}, \bibinfo {author} {\bibfnamefont {R.}~\bibnamefont {Otten}}, \ and\ \bibinfo {author} {\bibfnamefont {H.}~\bibnamefont {Bluhm}},\ }\href {\doibase 10.1103/physrevapplied.13.044071} {\bibfield  {journal} {\bibinfo  {journal} {Physical Review Applied}\ }\textbf {\bibinfo {volume} {13}},\ \bibinfo {pages} {044071} (\bibinfo {year} {2020})}\BibitemShut {NoStop}%
\bibitem [{\citenamefont {Jandura}\ \emph {et~al.}(2023)\citenamefont {Jandura}, \citenamefont {Thompson},\ and\ \citenamefont {Pupillo}}]{Jandura2023}%
  \BibitemOpen
  \bibfield  {author} {\bibinfo {author} {\bibfnamefont {S.}~\bibnamefont {Jandura}}, \bibinfo {author} {\bibfnamefont {J.~D.}\ \bibnamefont {Thompson}}, \ and\ \bibinfo {author} {\bibfnamefont {G.}~\bibnamefont {Pupillo}},\ }\href {\doibase 10.1103/prxquantum.4.020336} {\bibfield  {journal} {\bibinfo  {journal} {PRX Quantum}\ }\textbf {\bibinfo {volume} {4}},\ \bibinfo {pages} {020336} (\bibinfo {year} {2023})}\BibitemShut {NoStop}%
\bibitem [{Note1()}]{Note1}%
  \BibitemOpen
  \bibinfo {note} {Note that the 'control' of the CZZ gate is on the measurement qubit, but since $\protect \mathrm {CZZ} = \mathinner {|{0}\rangle }\mathinner {\langle {0}|} \otimes II + \mathinner {|{1}\rangle }\mathinner {\langle {1}|} \otimes ZZ = I \otimes (\mathinner {|{00}\rangle }\mathinner {\langle {00}|} +\mathinner {|{11}\rangle }\mathinner {\langle {11}|}) + Z \otimes (\mathinner {|{01}\rangle }\mathinner {\langle {01}|} +\mathinner {|{10}\rangle }\mathinner {\langle {10}|})$, this gate can also be thought of as a $X$-parity-controlled $Z$-gate.}\BibitemShut {Stop}%
\bibitem [{\citenamefont {Kribs}\ \emph {et~al.}(2005{\natexlab{a}})\citenamefont {Kribs}, \citenamefont {Laflamme},\ and\ \citenamefont {Poulin}}]{Kribs2005}%
  \BibitemOpen
  \bibfield  {author} {\bibinfo {author} {\bibfnamefont {D.}~\bibnamefont {Kribs}}, \bibinfo {author} {\bibfnamefont {R.}~\bibnamefont {Laflamme}}, \ and\ \bibinfo {author} {\bibfnamefont {D.}~\bibnamefont {Poulin}},\ }\href {\doibase 10.1103/physrevlett.94.180501} {\bibfield  {journal} {\bibinfo  {journal} {Physical Review Letters}\ }\textbf {\bibinfo {volume} {94}},\ \bibinfo {pages} {180501} (\bibinfo {year} {2005}{\natexlab{a}})}\BibitemShut {NoStop}%
\bibitem [{\citenamefont {Kribs}\ \emph {et~al.}(2005{\natexlab{b}})\citenamefont {Kribs}, \citenamefont {Laflamme}, \citenamefont {Poulin},\ and\ \citenamefont {Lesosky}}]{Kribs2005a}%
  \BibitemOpen
  \bibfield  {author} {\bibinfo {author} {\bibfnamefont {D.~W.}\ \bibnamefont {Kribs}}, \bibinfo {author} {\bibfnamefont {R.}~\bibnamefont {Laflamme}}, \bibinfo {author} {\bibfnamefont {D.}~\bibnamefont {Poulin}}, \ and\ \bibinfo {author} {\bibfnamefont {M.}~\bibnamefont {Lesosky}},\ }\href {\doibase 10.48550/ARXIV.QUANT-PH/0504189} {\  (\bibinfo {year} {2005}{\natexlab{b}}),\ 10.48550/ARXIV.QUANT-PH/0504189}\BibitemShut {NoStop}%
\bibitem [{\citenamefont {Hastings}\ and\ \citenamefont {Haah}(2021)}]{Hastings2021}%
  \BibitemOpen
  \bibfield  {author} {\bibinfo {author} {\bibfnamefont {M.~B.}\ \bibnamefont {Hastings}}\ and\ \bibinfo {author} {\bibfnamefont {J.}~\bibnamefont {Haah}},\ }\href {\doibase 10.22331/q-2021-10-19-564} {\bibfield  {journal} {\bibinfo  {journal} {Quantum}\ }\textbf {\bibinfo {volume} {5}},\ \bibinfo {pages} {564} (\bibinfo {year} {2021})}\BibitemShut {NoStop}%
\bibitem [{\citenamefont {Davydova}\ \emph {et~al.}(2023)\citenamefont {Davydova}, \citenamefont {Tantivasadakarn},\ and\ \citenamefont {Balasubramanian}}]{Davydova2023}%
  \BibitemOpen
  \bibfield  {author} {\bibinfo {author} {\bibfnamefont {M.}~\bibnamefont {Davydova}}, \bibinfo {author} {\bibfnamefont {N.}~\bibnamefont {Tantivasadakarn}}, \ and\ \bibinfo {author} {\bibfnamefont {S.}~\bibnamefont {Balasubramanian}},\ }\href {\doibase 10.1103/prxquantum.4.020341} {\bibfield  {journal} {\bibinfo  {journal} {PRX Quantum}\ }\textbf {\bibinfo {volume} {4}},\ \bibinfo {pages} {020341} (\bibinfo {year} {2023})}\BibitemShut {NoStop}%
\bibitem [{\citenamefont {Magdalena de~la Fuente}\ \emph {et~al.}(2025)\citenamefont {Magdalena de~la Fuente}, \citenamefont {Old}, \citenamefont {Townsend-Teague}, \citenamefont {Rispler}, \citenamefont {Eisert},\ and\ \citenamefont {Müller}}]{MagdalenadelaFuente2025}%
  \BibitemOpen
  \bibfield  {author} {\bibinfo {author} {\bibfnamefont {J.~C.}\ \bibnamefont {Magdalena de~la Fuente}}, \bibinfo {author} {\bibfnamefont {J.}~\bibnamefont {Old}}, \bibinfo {author} {\bibfnamefont {A.}~\bibnamefont {Townsend-Teague}}, \bibinfo {author} {\bibfnamefont {M.}~\bibnamefont {Rispler}}, \bibinfo {author} {\bibfnamefont {J.}~\bibnamefont {Eisert}}, \ and\ \bibinfo {author} {\bibfnamefont {M.}~\bibnamefont {Müller}},\ }\href {\doibase 10.1103/prxquantum.6.010360} {\bibfield  {journal} {\bibinfo  {journal} {PRX Quantum}\ }\textbf {\bibinfo {volume} {6}},\ \bibinfo {pages} {010360} (\bibinfo {year} {2025})}\BibitemShut {NoStop}%
\bibitem [{\citenamefont {Sung}\ \emph {et~al.}(2021)\citenamefont {Sung}, \citenamefont {Ding}, \citenamefont {Braumüller}, \citenamefont {Vepsäläinen}, \citenamefont {Kannan}, \citenamefont {Kjaergaard}, \citenamefont {Greene}, \citenamefont {Samach}, \citenamefont {McNally}, \citenamefont {Kim}, \citenamefont {Melville}, \citenamefont {Niedzielski}, \citenamefont {Schwartz}, \citenamefont {Yoder}, \citenamefont {Orlando}, \citenamefont {Gustavsson},\ and\ \citenamefont {Oliver}}]{Sung2021}%
  \BibitemOpen
  \bibfield  {author} {\bibinfo {author} {\bibfnamefont {Y.}~\bibnamefont {Sung}}, \bibinfo {author} {\bibfnamefont {L.}~\bibnamefont {Ding}}, \bibinfo {author} {\bibfnamefont {J.}~\bibnamefont {Braumüller}}, \bibinfo {author} {\bibfnamefont {A.}~\bibnamefont {Vepsäläinen}}, \bibinfo {author} {\bibfnamefont {B.}~\bibnamefont {Kannan}}, \bibinfo {author} {\bibfnamefont {M.}~\bibnamefont {Kjaergaard}}, \bibinfo {author} {\bibfnamefont {A.}~\bibnamefont {Greene}}, \bibinfo {author} {\bibfnamefont {G.~O.}\ \bibnamefont {Samach}}, \bibinfo {author} {\bibfnamefont {C.}~\bibnamefont {McNally}}, \bibinfo {author} {\bibfnamefont {D.}~\bibnamefont {Kim}}, \bibinfo {author} {\bibfnamefont {A.}~\bibnamefont {Melville}}, \bibinfo {author} {\bibfnamefont {B.~M.}\ \bibnamefont {Niedzielski}}, \bibinfo {author} {\bibfnamefont {M.~E.}\ \bibnamefont {Schwartz}}, \bibinfo {author} {\bibfnamefont {J.~L.}\ \bibnamefont {Yoder}}, \bibinfo {author} {\bibfnamefont {T.~P.}\ \bibnamefont {Orlando}}, \bibinfo {author} {\bibfnamefont
  {S.}~\bibnamefont {Gustavsson}}, \ and\ \bibinfo {author} {\bibfnamefont {W.~D.}\ \bibnamefont {Oliver}},\ }\href {\doibase 10.1103/physrevx.11.021058} {\bibfield  {journal} {\bibinfo  {journal} {Physical Review X}\ }\textbf {\bibinfo {volume} {11}},\ \bibinfo {pages} {021058} (\bibinfo {year} {2021})}\BibitemShut {NoStop}%
\bibitem [{\citenamefont {Collodo}\ \emph {et~al.}(2020)\citenamefont {Collodo}, \citenamefont {Herrmann}, \citenamefont {Lacroix}, \citenamefont {Andersen}, \citenamefont {Remm}, \citenamefont {Lazar}, \citenamefont {Besse}, \citenamefont {Walter}, \citenamefont {Wallraff},\ and\ \citenamefont {Eichler}}]{Collodo2020}%
  \BibitemOpen
  \bibfield  {author} {\bibinfo {author} {\bibfnamefont {M.~C.}\ \bibnamefont {Collodo}}, \bibinfo {author} {\bibfnamefont {J.}~\bibnamefont {Herrmann}}, \bibinfo {author} {\bibfnamefont {N.}~\bibnamefont {Lacroix}}, \bibinfo {author} {\bibfnamefont {C.~K.}\ \bibnamefont {Andersen}}, \bibinfo {author} {\bibfnamefont {A.}~\bibnamefont {Remm}}, \bibinfo {author} {\bibfnamefont {S.}~\bibnamefont {Lazar}}, \bibinfo {author} {\bibfnamefont {J.-C.}\ \bibnamefont {Besse}}, \bibinfo {author} {\bibfnamefont {T.}~\bibnamefont {Walter}}, \bibinfo {author} {\bibfnamefont {A.}~\bibnamefont {Wallraff}}, \ and\ \bibinfo {author} {\bibfnamefont {C.}~\bibnamefont {Eichler}},\ }\href {\doibase 10.1103/physrevlett.125.240502} {\bibfield  {journal} {\bibinfo  {journal} {Physical Review Letters}\ }\textbf {\bibinfo {volume} {125}},\ \bibinfo {pages} {240502} (\bibinfo {year} {2020})}\BibitemShut {NoStop}%
\bibitem [{\citenamefont {Heunisch}\ \emph {et~al.}(2023)\citenamefont {Heunisch}, \citenamefont {Eichler},\ and\ \citenamefont {Hartmann}}]{Heunisch2023}%
  \BibitemOpen
  \bibfield  {author} {\bibinfo {author} {\bibfnamefont {L.}~\bibnamefont {Heunisch}}, \bibinfo {author} {\bibfnamefont {C.}~\bibnamefont {Eichler}}, \ and\ \bibinfo {author} {\bibfnamefont {M.~J.}\ \bibnamefont {Hartmann}},\ }\href {\doibase 10.1103/physrevapplied.20.064037} {\bibfield  {journal} {\bibinfo  {journal} {Physical Review Applied}\ }\textbf {\bibinfo {volume} {20}},\ \bibinfo {pages} {064037} (\bibinfo {year} {2023})}\BibitemShut {NoStop}%
\bibitem [{\citenamefont {Vool}\ and\ \citenamefont {Devoret}(2017)}]{Vool2017}%
  \BibitemOpen
  \bibfield  {author} {\bibinfo {author} {\bibfnamefont {U.}~\bibnamefont {Vool}}\ and\ \bibinfo {author} {\bibfnamefont {M.}~\bibnamefont {Devoret}},\ }\href {\doibase 10.1002/cta.2359} {\bibfield  {journal} {\bibinfo  {journal} {International Journal of Circuit Theory and Applications}\ }\textbf {\bibinfo {volume} {45}},\ \bibinfo {pages} {897} (\bibinfo {year} {2017})}\BibitemShut {NoStop}%
\bibitem [{\citenamefont {Rasmussen}\ \emph {et~al.}(2021)\citenamefont {Rasmussen}, \citenamefont {Christensen}, \citenamefont {Pedersen}, \citenamefont {Kristensen}, \citenamefont {Bækkegaard}, \citenamefont {Loft},\ and\ \citenamefont {Zinner}}]{Rasmussen2021}%
  \BibitemOpen
  \bibfield  {author} {\bibinfo {author} {\bibfnamefont {S.}~\bibnamefont {Rasmussen}}, \bibinfo {author} {\bibfnamefont {K.}~\bibnamefont {Christensen}}, \bibinfo {author} {\bibfnamefont {S.}~\bibnamefont {Pedersen}}, \bibinfo {author} {\bibfnamefont {L.}~\bibnamefont {Kristensen}}, \bibinfo {author} {\bibfnamefont {T.}~\bibnamefont {Bækkegaard}}, \bibinfo {author} {\bibfnamefont {N.}~\bibnamefont {Loft}}, \ and\ \bibinfo {author} {\bibfnamefont {N.}~\bibnamefont {Zinner}},\ }\href {\doibase 10.1103/prxquantum.2.040204} {\bibfield  {journal} {\bibinfo  {journal} {PRX Quantum}\ }\textbf {\bibinfo {volume} {2}},\ \bibinfo {pages} {040204} (\bibinfo {year} {2021})}\BibitemShut {NoStop}%
\bibitem [{\citenamefont {Li}(2022)}]{Li2022}%
  \BibitemOpen
  \bibfield  {author} {\bibinfo {author} {\bibfnamefont {S.~e.~a.}\ \bibnamefont {Li}},\ }\href {\doibase 10.1088/0256-307x/39/3/030302} {\bibfield  {journal} {\bibinfo  {journal} {Chinese Physics Letters}\ }\textbf {\bibinfo {volume} {39}},\ \bibinfo {pages} {030302} (\bibinfo {year} {2022})}\BibitemShut {NoStop}%
\bibitem [{\citenamefont {Baker}\ \emph {et~al.}(2022)\citenamefont {Baker}, \citenamefont {Huber}, \citenamefont {Glaser}, \citenamefont {Roy}, \citenamefont {Tsitsilin}, \citenamefont {Filipp},\ and\ \citenamefont {Hartmann}}]{Baker2022}%
  \BibitemOpen
  \bibfield  {author} {\bibinfo {author} {\bibfnamefont {A.~J.}\ \bibnamefont {Baker}}, \bibinfo {author} {\bibfnamefont {G.~B.~P.}\ \bibnamefont {Huber}}, \bibinfo {author} {\bibfnamefont {N.~J.}\ \bibnamefont {Glaser}}, \bibinfo {author} {\bibfnamefont {F.}~\bibnamefont {Roy}}, \bibinfo {author} {\bibfnamefont {I.}~\bibnamefont {Tsitsilin}}, \bibinfo {author} {\bibfnamefont {S.}~\bibnamefont {Filipp}}, \ and\ \bibinfo {author} {\bibfnamefont {M.~J.}\ \bibnamefont {Hartmann}},\ }\href {\doibase 10.1063/5.0077443} {\bibfield  {journal} {\bibinfo  {journal} {Applied Physics Letters}\ }\textbf {\bibinfo {volume} {120}} (\bibinfo {year} {2022}),\ 10.1063/5.0077443}\BibitemShut {NoStop}%
\bibitem [{\citenamefont {Wang}(2022)}]{Wang2022}%
  \BibitemOpen
  \bibfield  {author} {\bibinfo {author} {\bibfnamefont {C.~e.~a.}\ \bibnamefont {Wang}},\ }\href {\doibase 10.1038/s41534-021-00510-2} {\bibfield  {journal} {\bibinfo  {journal} {npj Quantum Information}\ }\textbf {\bibinfo {volume} {8}} (\bibinfo {year} {2022}),\ 10.1038/s41534-021-00510-2}\BibitemShut {NoStop}%
\bibitem [{\citenamefont {Perlin}(2023)}]{Perlin2023}%
  \BibitemOpen
  \bibfield  {author} {\bibinfo {author} {\bibfnamefont {M.~A.}\ \bibnamefont {Perlin}},\ }\href {\doibase 10.48550/ARXIV.2311.09129} {\enquote {\bibinfo {title} {A short note on effective pauli noise models},}\ } (\bibinfo {year} {2023})\BibitemShut {NoStop}%
\bibitem [{\citenamefont {Geller}\ and\ \citenamefont {Zhou}(2013)}]{Geller2013}%
  \BibitemOpen
  \bibfield  {author} {\bibinfo {author} {\bibfnamefont {M.~R.}\ \bibnamefont {Geller}}\ and\ \bibinfo {author} {\bibfnamefont {Z.}~\bibnamefont {Zhou}},\ }\href {\doibase 10.1103/physreva.88.012314} {\bibfield  {journal} {\bibinfo  {journal} {Physical Review A}\ }\textbf {\bibinfo {volume} {88}},\ \bibinfo {pages} {012314} (\bibinfo {year} {2013})}\BibitemShut {NoStop}%
\bibitem [{\citenamefont {McEwen}(2021)}]{McEwen2021}%
  \BibitemOpen
  \bibfield  {author} {\bibinfo {author} {\bibfnamefont {M.~e.~a.}\ \bibnamefont {McEwen}},\ }\href {\doibase 10.1038/s41467-021-21982-y} {\bibfield  {journal} {\bibinfo  {journal} {Nature Communications}\ }\textbf {\bibinfo {volume} {12}} (\bibinfo {year} {2021}),\ 10.1038/s41467-021-21982-y}\BibitemShut {NoStop}%
\bibitem [{\citenamefont {Miao}(2023)}]{Miao2023}%
  \BibitemOpen
  \bibfield  {author} {\bibinfo {author} {\bibfnamefont {K.~C. e.~a.}\ \bibnamefont {Miao}},\ }\href {\doibase 10.1038/s41567-023-02226-w} {\bibfield  {journal} {\bibinfo  {journal} {Nature Physics}\ }\textbf {\bibinfo {volume} {19}},\ \bibinfo {pages} {1780} (\bibinfo {year} {2023})}\BibitemShut {NoStop}%
\bibitem [{\citenamefont {Jandura}\ and\ \citenamefont {Pupillo}(2024)}]{Jandura2024}%
  \BibitemOpen
  \bibfield  {author} {\bibinfo {author} {\bibfnamefont {S.}~\bibnamefont {Jandura}}\ and\ \bibinfo {author} {\bibfnamefont {G.}~\bibnamefont {Pupillo}},\ }\href {\doibase 10.48550/ARXIV.2405.16621} {\enquote {\bibinfo {title} {Surface code stabilizer measurements for rydberg atoms},}\ } (\bibinfo {year} {2024})\BibitemShut {NoStop}%
\bibitem [{\citenamefont {Lacroix}\ \emph {et~al.}(2025)\citenamefont {Lacroix}, \citenamefont {Hofele}, \citenamefont {Remm}, \citenamefont {Benhayoune-Khadraoui}, \citenamefont {McDonald}, \citenamefont {Shillito}, \citenamefont {Lazar}, \citenamefont {Hellings}, \citenamefont {Swiadek}, \citenamefont {Colao-Zanuz}, \citenamefont {Flasby}, \citenamefont {Panah}, \citenamefont {Kerschbaum}, \citenamefont {Norris}, \citenamefont {Blais}, \citenamefont {Wallraff},\ and\ \citenamefont {Krinner}}]{Lacroix2025}%
  \BibitemOpen
  \bibfield  {author} {\bibinfo {author} {\bibfnamefont {N.}~\bibnamefont {Lacroix}}, \bibinfo {author} {\bibfnamefont {L.}~\bibnamefont {Hofele}}, \bibinfo {author} {\bibfnamefont {A.}~\bibnamefont {Remm}}, \bibinfo {author} {\bibfnamefont {O.}~\bibnamefont {Benhayoune-Khadraoui}}, \bibinfo {author} {\bibfnamefont {A.}~\bibnamefont {McDonald}}, \bibinfo {author} {\bibfnamefont {R.}~\bibnamefont {Shillito}}, \bibinfo {author} {\bibfnamefont {S.}~\bibnamefont {Lazar}}, \bibinfo {author} {\bibfnamefont {C.}~\bibnamefont {Hellings}}, \bibinfo {author} {\bibfnamefont {F.}~\bibnamefont {Swiadek}}, \bibinfo {author} {\bibfnamefont {D.}~\bibnamefont {Colao-Zanuz}}, \bibinfo {author} {\bibfnamefont {A.}~\bibnamefont {Flasby}}, \bibinfo {author} {\bibfnamefont {M.~B.}\ \bibnamefont {Panah}}, \bibinfo {author} {\bibfnamefont {M.}~\bibnamefont {Kerschbaum}}, \bibinfo {author} {\bibfnamefont {G.~J.}\ \bibnamefont {Norris}}, \bibinfo {author} {\bibfnamefont {A.}~\bibnamefont {Blais}}, \bibinfo {author} {\bibfnamefont
  {A.}~\bibnamefont {Wallraff}}, \ and\ \bibinfo {author} {\bibfnamefont {S.}~\bibnamefont {Krinner}},\ }\href {\doibase 10.1103/physrevlett.134.120601} {\bibfield  {journal} {\bibinfo  {journal} {Physical Review Letters}\ }\textbf {\bibinfo {volume} {134}},\ \bibinfo {pages} {120601} (\bibinfo {year} {2025})}\BibitemShut {NoStop}%
\bibitem [{\citenamefont {Old}\ \emph {et~al.}(2025)\citenamefont {Old}, \citenamefont {Tasler}, \citenamefont {Hartmann},\ and\ \citenamefont {Müller}}]{old2025faulttolerant}%
  \BibitemOpen
  \bibfield  {author} {\bibinfo {author} {\bibfnamefont {J.}~\bibnamefont {Old}}, \bibinfo {author} {\bibfnamefont {S.}~\bibnamefont {Tasler}}, \bibinfo {author} {\bibfnamefont {M.~J.}\ \bibnamefont {Hartmann}}, \ and\ \bibinfo {author} {\bibfnamefont {M.}~\bibnamefont {Müller}},\ }\href@noop {} {\bibfield  {journal} {\bibinfo  {journal} {to appear}\ } (\bibinfo {year} {2025})}\BibitemShut {NoStop}%
\bibitem [{\citenamefont {Gidney}(2021)}]{Gidney2021}%
  \BibitemOpen
  \bibfield  {author} {\bibinfo {author} {\bibfnamefont {C.}~\bibnamefont {Gidney}},\ }\href {\doibase 10.22331/q-2021-07-06-497} {\bibfield  {journal} {\bibinfo  {journal} {Quantum}\ }\textbf {\bibinfo {volume} {5}},\ \bibinfo {pages} {497} (\bibinfo {year} {2021})}\BibitemShut {NoStop}%
\bibitem [{\citenamefont {Johansson}\ \emph {et~al.}(2012)\citenamefont {Johansson}, \citenamefont {Nation},\ and\ \citenamefont {Nori}}]{Johansson2012}%
  \BibitemOpen
  \bibfield  {author} {\bibinfo {author} {\bibfnamefont {J.}~\bibnamefont {Johansson}}, \bibinfo {author} {\bibfnamefont {P.}~\bibnamefont {Nation}}, \ and\ \bibinfo {author} {\bibfnamefont {F.}~\bibnamefont {Nori}},\ }\href {\doibase 10.1016/j.cpc.2012.02.021} {\bibfield  {journal} {\bibinfo  {journal} {Computer Physics Communications}\ }\textbf {\bibinfo {volume} {183}},\ \bibinfo {pages} {1760} (\bibinfo {year} {2012})}\BibitemShut {NoStop}%
\bibitem [{\citenamefont {Johansson}\ \emph {et~al.}(2013)\citenamefont {Johansson}, \citenamefont {Nation},\ and\ \citenamefont {Nori}}]{Johansson2013}%
  \BibitemOpen
  \bibfield  {author} {\bibinfo {author} {\bibfnamefont {J.}~\bibnamefont {Johansson}}, \bibinfo {author} {\bibfnamefont {P.}~\bibnamefont {Nation}}, \ and\ \bibinfo {author} {\bibfnamefont {F.}~\bibnamefont {Nori}},\ }\href {\doibase 10.1016/j.cpc.2012.11.019} {\bibfield  {journal} {\bibinfo  {journal} {Computer Physics Communications}\ }\textbf {\bibinfo {volume} {184}},\ \bibinfo {pages} {1234} (\bibinfo {year} {2013})}\BibitemShut {NoStop}%
\bibitem [{\citenamefont {Manes}\ and\ \citenamefont {Claes}(2025)}]{Manes2025}%
  \BibitemOpen
  \bibfield  {author} {\bibinfo {author} {\bibfnamefont {A.~G.}\ \bibnamefont {Manes}}\ and\ \bibinfo {author} {\bibfnamefont {J.}~\bibnamefont {Claes}},\ }\href {\doibase 10.22331/q-2025-01-30-1618} {\bibfield  {journal} {\bibinfo  {journal} {Quantum}\ }\textbf {\bibinfo {volume} {9}},\ \bibinfo {pages} {1618} (\bibinfo {year} {2025})}\BibitemShut {NoStop}%
\bibitem [{\citenamefont {Derks}\ \emph {et~al.}(2024)\citenamefont {Derks}, \citenamefont {Townsend-Teague}, \citenamefont {Burchards},\ and\ \citenamefont {Eisert}}]{Derks2024}%
  \BibitemOpen
  \bibfield  {author} {\bibinfo {author} {\bibfnamefont {P.-J. H.~S.}\ \bibnamefont {Derks}}, \bibinfo {author} {\bibfnamefont {A.}~\bibnamefont {Townsend-Teague}}, \bibinfo {author} {\bibfnamefont {A.~G.}\ \bibnamefont {Burchards}}, \ and\ \bibinfo {author} {\bibfnamefont {J.}~\bibnamefont {Eisert}},\ }\href {\doibase 10.48550/ARXIV.2407.13826} {\enquote {\bibinfo {title} {Designing fault-tolerant circuits using detector error models},}\ } (\bibinfo {year} {2024})\BibitemShut {NoStop}%
\bibitem [{\citenamefont {Higgott}(2021)}]{Higgott2021a}%
  \BibitemOpen
  \bibfield  {author} {\bibinfo {author} {\bibfnamefont {O.}~\bibnamefont {Higgott}},\ }\href {\doibase 10.48550/ARXIV.2105.13082} {\enquote {\bibinfo {title} {Pymatching: A python package for decoding quantum codes with minimum-weight perfect matching},}\ } (\bibinfo {year} {2021})\BibitemShut {NoStop}%
\bibitem [{\citenamefont {Sorge}(2015)}]{Sorge2015}%
  \BibitemOpen
  \bibfield  {author} {\bibinfo {author} {\bibfnamefont {A.}~\bibnamefont {Sorge}},\ }\href {\doibase 10.5281/zenodo.35293} {\  (\bibinfo {year} {2015}),\ 10.5281/zenodo.35293}\BibitemShut {NoStop}%
\end{thebibliography}%

\appendix
\onecolumngrid
\section{Circuit quantisation} \label{apx:circuit_quantisation}
\begin{figure}[h]
    \vspace{0.3cm}
    \centering
    \includegraphics[width=0.5\textwidth]{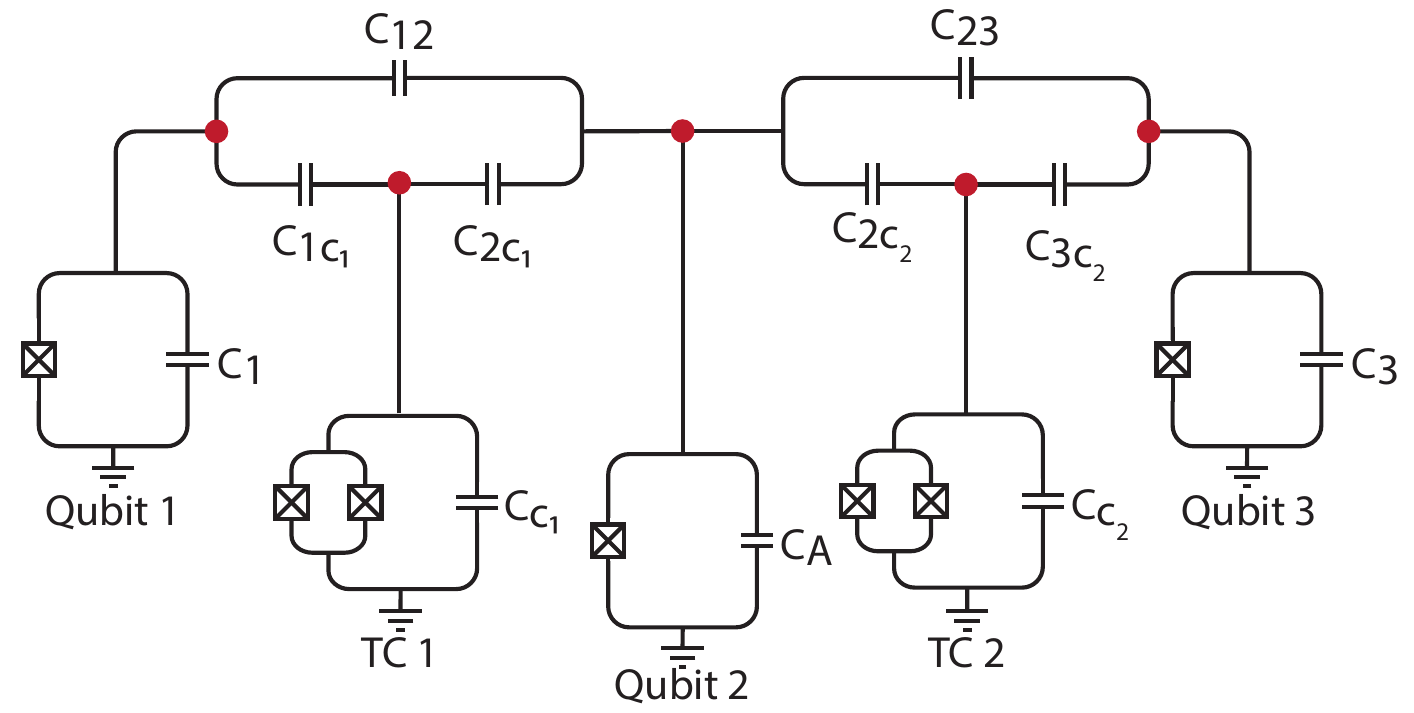}
    \caption{Circuit diagram of the parity check circuit. The active nodes are depicted by the red dots.}
    \label{circuit_diagram_nodes}
\end{figure}

The circuit given in Fig.\,\ref{circuit_diagram} is quantized in the standard QED procedure \cite{Vool2017,Rasmussen2021}, therefore the active nodes are determined by choosing a spanning tree. The active nodes are depicted in Fig.\,\ref{circuit_diagram_nodes} by the red dots. Thus the active node fluxes can be written in a flux vector 
\begin{equation}
    \phi = \begin{pmatrix}
    \phi_1 \\
    \phi_{c1} \\
    \phi_2\\
    \phi_{c2}\\
    \phi_3
    \end{pmatrix}.
\end{equation}
With the flux vector the capacitance matrix \textbf{C} can be set up.
\begin{equation}
      C = \begin{pmatrix}
C_{1c_1} + C_{12} + C_1 & - C_{1c_1} & - C_{12} & 0 & 0 \\
- C_{1c_1} & C_{1c_1} + C_{2c_1} + C_{c_1} & - C_{2c_1} & 0 & 0 \\
- C_{12} & - C_{2c_1} & C_{12} + C_{2c_1} + C_{2c_2} + C_{23} + C_2 & -C_{2c_2} & -C_{23} \\
0 & 0 & -C_{2c_2} & C_{2c_2} + C_{3c_2} + C_{c_2} & -C_{3c_2} \\
0 & 0 & -C_{23} & -C_{3c_2} & C_{23} + C_{3c_2} + C_3
\end{pmatrix}
\end{equation}
The Lagrangian is given by 
\begin{equation}
    \mathcal{L} = \frac{1}{2}\dot{\bm{\phi}}^{\text{ T}} \bm{C} \dot{\bm{\phi}} - E_{pot}.
\end{equation}
The capacitance matrix can now be used to calculate the conjugate charges for each node by $\bm{q} = \bm{C} \dot{\bm{\phi}}$. The Hamiltonian can be derived by applying a Legendre transformation on the Lagrangian. For this one needs to express $\bm{\dot{\phi}}$ in terms of $\bm{q}$. This is done by inverting the capacitance matrix.
\begin{equation}
    \mathcal{H} = \dot{\bm{\phi}}^T\bm{q} - \mathcal{L} = \frac{1}{2} \bm{q}^T \bm{C}^{-1} \bm{q} - E_{\text{pot}}(\bm{\phi})
\end{equation}
Furthermore the effective capacitative energy matrix $\bm{E}_{\text{C}_{\text{ij}}}$ can be defined as
\begin{equation}
    \bm{E}_{\text{C}_{\text{ij}}} = \frac{e^2}{2} \bm{C}^{-1}_{ij}.
    \label{EC_matrix}
\end{equation}
The net number of Cooper pairs for the mode i can be written as
\begin{equation}
    \hat{n}_i = -\frac{\hat{q}_i}{2e}.
    \label{CP_number}
\end{equation}
And the potential energy of the Josephson Junctions is defined as
\begin{align*}
    E_{\text{pot}} = 
    &-E_{J_1}\text{cos}(\phi_1)
    -E_{J_{\text{c}_1}}\text{cos}(\phi_{\text{c}_1}) \text{cos}(\varphi_{c_1}^{ext}) \\
    &-E_{J_2}\text{cos}(\phi_{2})
    -E_{J_{\text{c}_2}}\text{cos}(\phi_{\text{c}_2}) \text{cos}(\varphi_{c_2}^{ext})\\
    &-E_{J_3}\text{cos}(\phi_3). \numberthis \label{JJ_energy_2QA}
\end{align*}
After quantization, equations \ref{EC_matrix}, \ref{CP_number}, \ref{JJ_energy_2QA} can now be used to rewrite the system Hamiltonian in terms of the quantized Cooper pair number $\hat{n}$ and the quantized flux $\hat{\phi}$. In the following, the transmon approximation of the potential energy $E_{J} \, \text{cos}(\phi) \simeq E_{J} - \frac{1}{2}E_{J} \phi^2 + \frac{1}{24}E_{J} \phi^4$ is used. For clarity reasons, in the following part the hat is left out for quantum mechanical operators. 
\begin{align*}
    H &= \sum_{i= 1,\text{c}_1,2,\text{c}_2,3} 4\ E_{\text{C}_{ii}} \ n_i^2 + \frac{1}{2}E_{J_i} \phi_i^2 -\frac{1}{24} E_{J_i} \phi_i^4 \\
    &+ \sum_{k=1,2,3; l=1,2} 8E_{\text{C}_{kc_l}} n_k n_{c_l} + \sum_{m<o=1,2,3} 8E_{\text{C}_{mo}} n_m n_o \\
    &+ 8E_{\text{C}_{c_1c_2}} n_{c_1} n_{c_2}  \numberthis
\end{align*}
The obtained Hamiltonian can now be written in second quantization by introducing creation and annihilation operator $b_i^\dag, b_i$, where $i=1,A,3$ for the data qubits and ancilla qubit and $c_k^\dag, c_k$, where $k=1,2$ for the two couplers. The Hamiltonian in second quantisation has now the following shape
\begin{align*}
    H&= \sum_{i=1,2,3}\omega_ib_i^\dagger b_i +\frac{\alpha_i}{2}b_i^\dagger b_i^\dagger b_i b_i \\ 
    &+\sum_{j=1,2}\omega_{\text{c}_j}c_j^\dagger c_j +\frac{\alpha_j}{2}c_j^\dagger c_j^\dagger c_j c_j \\
    &+\sum_{k=1,2,3; l=1,2} g_{k,c_l}(b_k - b_k^\dagger) (c_l - c_l^\dagger)\\
    &+\sum_{m<o=1,2,3} g_{mo} (b_m - b_m^\dagger) (b_o - b_o^\dagger)\\
    &+ g_{c_1c_2}(c_1 - c_1^\dagger) (c_2 - c_2^\dagger) \numberthis \label{apx:2QA_circuit_Hamiltonian}
\end{align*}

\section{Parameter Optimization} \label{apx:parameter_optimization}
For obtaining an optimal parity check gate, the main challenge is to find adequate circuit and pulse parameters. Since the parity check gate should produce the same $ZZ$-interaction between each data qubit and ancilla qubit, it is beneficial to choose symmetric parameters for the circuit model given in Fig.\,\ref{circuit_diagram}. For the choice of capacitance values and also frequency values of the fixed frequency qubits, \cite{Collodo2020, Baker2022} were used as orientation. The final values for the fixed circuit parameter are displayed in Table \ref{tab:2QA_capacitances} and \ref{tab:2QA_frequencies}.

\begin{table}[]
    \centering
    \begin{tabular}{|Sc|Sc|}
    \hline
        Capacitor & Capacity in fF \\ \Xhline{1pt}
        $C_1 = C_2 = C_3 $ & $77.8$  \\ \hline
         $C_{c_1} = C_{c_2}$ & $60.4$ \\ \hline
         $C_{12} = C_{23}$ & $0.46$ \\ \hline
         $C_{1c_1} = C_{2c_2} = C_{2c_1} = C_{2c_2} $ & $6.4$ \\ \hline
    \end{tabular}
    \caption{The Table shows the chosen capacitor values for the two qubit ancilla circuit. The capacitances were chosen symmetrically to obtain similar $ZZ$-couplings between data qubits and ancilla.}
    \label{tab:2QA_capacitances}
\end{table}
\begin{table}[]
    \centering
    \begin{tabular}{|Sc|Sc|}
    \hline
        \multirow{2}*{\ Qubit} & Frequency \ \\[-0.5em]
        &[GHz] \ \\ \Xhline{1pt}
         $\omega_1$ & $4.89$ \\ \hline
         $\omega_2$ & $5.31$ \\ \hline
         $\omega_3$ & $4.83$ \\ \hline
    \end{tabular}
    \caption{The Table shows the frequency choices for the three different fixed frequency qubits Q1, Q2 and Q3.}
    \label{tab:2QA_frequencies}
\end{table}

After fixing the capacitances and fixed qubit frequencies, the dynamical circuit parameters, the frequencies of the two tunable couplers, are obtained by using optimal control. In the idle position the two couplers are parked at the frequencies $\omega_{\text{c}_1} = 7.496\,$GHz and $\omega_{\text{c}_2} = 7.44\,$GHz. To determine the pulses, which tune the two couplers, an optimal control optimization is performed. The pulse function is based on a flattop Gaussian function, which has the following shape.
\begin{equation}
    \begin{gathered}
        \frac{A_{c_i}}{4} \cdot \left( 1 + \text{erf} \left( \frac{t- \Delta t_i - t^i_{\text{ramp}}}{t^i_{\text{ramp}}} \right) \right) \cdot \left( 1 + \text{erf} \left( \frac{t^i_{\text{gate}}-t+ \Delta t_i -t^i_{\text{ramp}}}{t^i_{\text{ramp}}} \right) \right), \ \ i = 1,2
    \end{gathered}
\end{equation}
The optimizable parameters are the amplitude $\text{A}_{\text{c}_i}$, the gate time $t_{\text{gate}}^{i}$ and the ramp up time $t_{\text{ramp}}^{i}$, the index $i$ represents the corresponding coupler and runs from 1 to 2. In contrast to common gate optimizations we use a cost function given in Eqn. \ref{eqn:new_cost}, that minimizes fault-tolerance breaking errors and leakage. For the calculation of the Pauli coefficients, we explained in section \ref{subsec:error_model}, we define the ideal CZZ unitary in the three qubit two level subspace consisting of the two data qubits and the ancilla qubit. We define the two data qubits ($\ket{i},\ket{j} \in (\ket{1},\ket{0})$) and the ancilla qubit ($\ket{k} \in (\ket{0}, \ket{1})$ in the $Z$-basis. This results in the 8x8 matrix mentioned in equation \ref{eqn:ideal_matrix}, which is constructed by the product states $\ket{ijk}$.

\begin{figure}
    \centering
    \includegraphics[width=\textwidth]{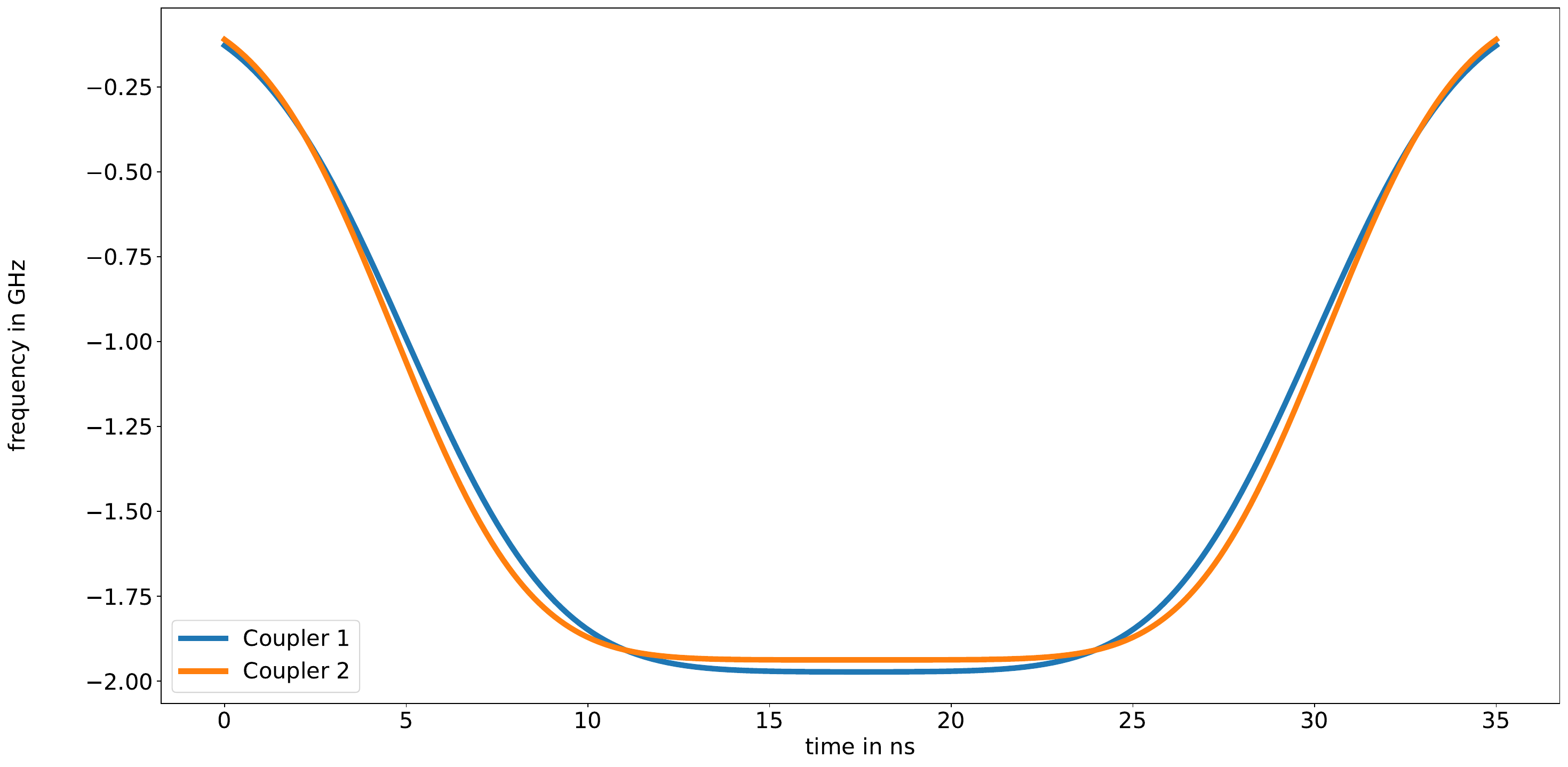}
    \caption{The two tunable couplers are strongly tuned close to the ancilla qubit frequency, this results in a large $ZZ$-coupling, which enables a fast parity check gate.}
    \label{pulse_signal}
\end{figure}

For the calculation of the Pauli weights, we also need the simulated Pauli channel. To obtain the channel $\mathcal{U}_{\text{sim}}$ from the simulation, the inner product of the input state and the state after the time evolution is taken, which can be written as
\begin{equation}
    \mathcal{U}_{\text{sim}}[i,j] = \braket{i|\mathcal{U}_{\text{gate}}|j}.
\end{equation}
where
$\ket{i},\ket{j} \in \{ \ket{000}, \ket{001}, \ket{010}, \ket{011}, \ket{100}, \ket{101},$ $\ket{110}, \ket{111} \}$
are representing the entries of the channel $\mathcal{U}_{\text{sim}}$. We use the python package QuTiP \cite{Johansson2012,Johansson2013} for the gate simulation, and the Nelder-Mead algorithm of SciPy minimize for the the pulse-optimization. The such obtained optimal pulses for the two couplers can be seen in Fig.\,\ref{pulse_signal}. We performed the full simulation of the parity check gate using QuTiP \cite{Johansson2012,Johansson2013}. 

\section{QEC simulations - numerical methods}

\subsection{Memory experiments} \label{apx:memory_exp}
We benchmark the performance of the error correction circuits in $Z$- ($X$-)\emph{memory experiments}. For a distance $d$ surface code, we
\begin{enumerate}
    \item Prepare $n$ data qubits in $\ket{0}^{\otimes n}$ ($\ket{+}^{\otimes n}$).
    \item Using $n-1$ ancilla qubits, perform $d$ rounds of ($Z$- and $X$-) syndrome measurements.
    \item Measure the $n$ data qubits in the $Z$- ($X$-) basis.
\end{enumerate}

We construct the circuits as described in the next section in \texttt{stim}\,\cite{Gidney2021} and  declare \texttt{DETECTOR}s representing sets of deterministic measurement as the parity of consecutive syndrome measurements. 
For a $Z$- ($X$-)basis memory experiment, the first $Z$-($X$-)stabilizer measurement is already deterministic and considered a detector on its own. Also, the last stabilizer measurement is compared to the syndrome reconstructed from the single qubit measurements. 
From the final single-qubit $Z$-($X$-) measurements, we can also reconstruct the value of the logical $Z$ ($X$) operator, which we annotate as \texttt{OBSERVABLE\_INCLUDE}.
Asymmetries arising from effective or biased noise channels, as well as stabilizer readout circuit details can influence the symmetry of $X$- and $Z$- memory experiments. We therefore always simulate both memory experiments and report 
\begin{align}
    p_L = 1 - (1-p_L^{(X)}) (1-p_L^{(Z)}) \label{eqn:logical_error_rate}
\end{align}
as the overall logical error rate assuming independence of $X$- and $Z$- logical error rates.

\subsection{Stabilizer measurement circuits}
\begin{figure}
    \centering
    \includegraphics[width=\linewidth]{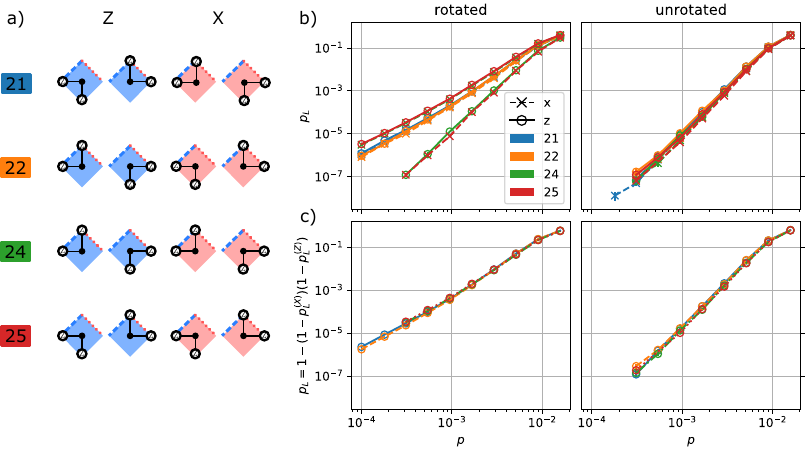}
    \caption{a) Orderings $21,22,24$ and $25$ of three-qubit CZZ gates for $Z$- and $X$-stabilizer measurements. On the plaquettes, we draw in dashed (dotted) lines the overlap of a $Z$- ($X$-) logical operator in the rotated surface code. b) Logical error rate of $X$- and $Z$-basis memory experiments on distance $d=7$ rotated surface codes. For orderings 21 and 22,  these show a symmetric behavior. Orderings 24 and 25 perform differently for the two bases as described in the text. c) Combining the $X$- and $Z$- logical error rates shows that the logical error rates are very similar across the orderings. }
    \label{fig:cz_directions}
\end{figure}

There are numerous ways to schedule and order the entangling gates required for the projective measurement of the stabilizer generators. 
While in unrotated surface codes, the order does not influence the fault-distance of the circuits\,\cite{Dennis2002,Manes2025}, rotated surface code require a scheduling such that the last two gates interact with data qubits that are \emph{orthogonal} to the logical operator corresponding to the Pauli type of the measured stabilizer\,\cite{Fowler2012}. 
We show the different schedules in Fig.\,\ref{fig:cz_directions} a), and the respective logical $X$- and $Z$- error rates for a distance $7$ surface codes implementing the CZZ gate using these schedule in Fig.\,\ref{fig:cz_directions} b). We use a uniform depolarizing noise model as described in the main text. These results confirm the ordering-independence of the unrotated surface code circuits. For the rotated surface codes and an ordering that is orthogonal (21) or parallel (22) to the respective logical, the logical error rates are also symmetric in $X$ and $Z$. We can, however, turn one of the memory experiments fault-tolerant by ordering both, $Z$- and $X$- stabilizer measurements orthogonal to a fixed logical operators. For ordering $24$, e.g., all gates are orthogonal to the $X$-logical. This allows for a distance preserving protection against $X$-logical errors, such that the $Z$-basis memory experiments shows the expected FT scaling (here $\propto p^4$ for $d=7$). This holds analogously for ordering $25$ with Paulis $X$ and $Z$ interchanged. In Fig.\,\ref{fig:cz_directions} c), we show the combined logical error rate (Eqn.\,\ref{eqn:logical_error_rate}). The asymmetry for different bases and orderings disappear when relying on this metric.

\begin{figure*}
    \centering
    \includegraphics[width=\linewidth]{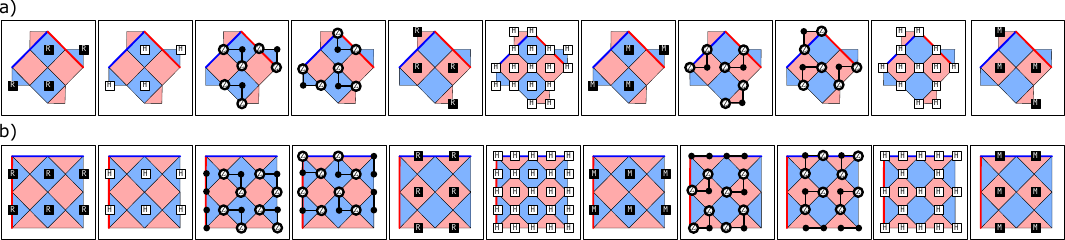}
    \caption{One round of stabilizer measurements of distance $3$ surface codes in the a) rotated and b) unrotated implementation using ordering $21$. $X$- and $Z$-stabilizers are drawn in light red and blue respectively. We also indicate minimum weight logical operators. CZZ gates always have their control qubit on the ancillary qubit in the center of a plaquette. The implementation with CZZ gates requires two timesteps of entangling gates per Pauli type, compared to $4$ timesteps for an equivalent implementation with CZ gates. }
    \label{fig:stab_readouts}
\end{figure*}

Finally, we show one round of stabilizer measurements using CZZ gates for rotated (a) and unrotated (b) surface codes in Fig.\,\ref{fig:stab_readouts}. Here, and in all simulations shown in the main text, we use ordering 21. In the actual simulations, we additionally reduce the number of idling locations by placing reset operations parallel to entangling gates (i.e.~merging timesteps $4$ and $5$ in Fig.\,\ref{fig:stab_readouts}).

\subsection{Noise Models}
If not otherwise specified, we implement a single-parameter, superconducting qubit architecture inspired \emph{circuit level} noise model with noise parameters 
\begin{align}
    p_{\mathrm{H}} = p_{\mathrm{CZ}}  = p_{\mathrm{CZZ}} &= p\\
    p_{\mathrm{idle}} &= 0.1p\\
    p_{\mathrm{reset}} &= 2p\\
    p_{\mathrm{measure}} &= 5p.
\end{align}

\subsection{Implementing three-qubit gates and $n$-qubit depolarizing errors in \texttt{stim}}
There is no native way to include $\mathrm{CZZ}$-gates and depolarizing channels acting on $n > 2$ qubits in \texttt{stim}.
We therefore place two $\mathrm{CZ}$-gates in the same \texttt{TICK} and use \texttt{stim}s included correlated error feature to mimic three-qubit depolarizing error channels. 
In the simulation, an \texttt{ELSE\_CORRELATED\_ERROR(p) P1*P2*\dots} only occurs with probability $p$ if none of the preceding \texttt{ELSE\_CORRELATED\_ERROR}s occurred. Given a non-uniform depolarizing channel on $n$ qubits with probabilities $\vec{p} = \{p_i\}_{i=1}^{4^n-1}$ of (Pauli) errors $\vec{E} = \{P_i\}_{i=1}^{4^n-1}$, we can construct the corresponding correlated error channel in \texttt{stim} by rescaling
\begin{align}
    \tilde{p}_i = \frac{p_i}{\prod_{j=0}^{i-1} (1-\tilde{p}_j)} = \frac{p_i}{1 - \sum_{j=0}^{i-1} p_j}.
\end{align}
As an example, consider a $2$-qubit bit-flip channel with probabilities $p(XI)=p_{XI}, p(IX)=p_{IX}, p(XX)=0$ and $p(II) = 1-p_{XI}-p_{IX}$. We cannot reconstruct the probability distribution of this channel using \emph{independent} single qubit channels, because then $p(XX) = p_{IX}p_{XI} + p_{XX} \neq 0$. Using correlated errors, this $2$-qubit bit-flip channel can be written as
\begin{align}
    \texttt{CORRELATED\_ERROR(p\_XI) X1}\phantom{.} \nonumber \\
    \texttt{ELSE\_CORRELATED\_ERROR(p\_IX/(1-p\_XI)) X2}.
\end{align}
For $p_{IX} = p_{XI} = 0.01$, this is
\begin{align}
    \texttt{CORRELATED\_ERROR(0.01) X1}\phantom{.} \nonumber \\
    \texttt{ELSE\_CORRELATED\_ERROR(0.01010101) X2}. 
\end{align}

\subsection{Decoding}
From the noisy circuit, we construct a \emph{detector error model}\,\cite{Derks2024} that is a list of independent error mechanisms and the corresponding detectors and observables that are flipped by them. From this list, we can construct a parity check matrix or decoding graph used to configure a decoder like pymatching\,\cite{Higgott2021a}. 

Note the \emph{independence} of error mechanisms in the detector error model, while the effective Pauli channels give non-independent errors on the $\mathrm{CZZ}$ gate. Sampling of the noisy circuit is done with the correlated errors, decoding, however, is done on a detector error model that approximates disjoint error channels as independent. 
This is handled automatically in \texttt{stim} by setting the flag '\texttt{approximate\_disjoint\_errors=True}'. 

\begin{figure}
    \centering
    \includegraphics[width=0.8\linewidth]{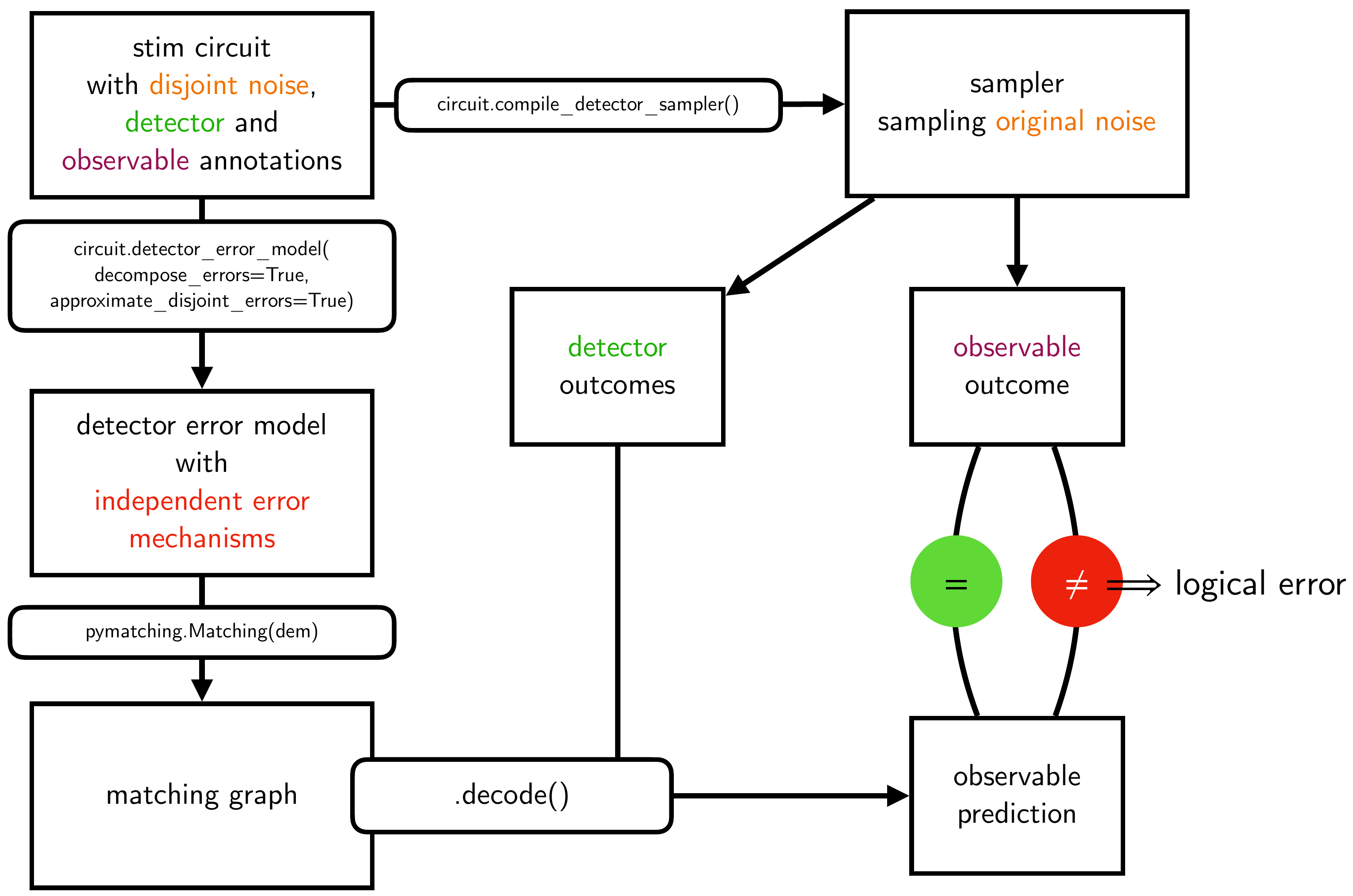}
    \caption{Simulation workflow using \texttt{stim} and \texttt{pymatching}. The circuit is converted to a detector error model that enables fast decoding using a matching graph. While this relies on independent error mechanisms and decomposes them into a matchable problem, we compile a detector sampler directly from the circuit to sample the circuits detector and observable outcomes under the influence of the actual noise. Whenever the observable prediction of the decoder is different from the actual simulated outcome, we record a logical error. }
    \label{fig:simulation_methods}
\end{figure}

A run of a memory experiment gives the values of detectors, the \emph{syndrome}, and values of observables, the \emph{logical outcome}. The syndrome is then fed to the decoder - which returns a proposal of error mechanisms flipped that produce the same syndrome. These error mechanisms also produce values of observables, the \emph{logical predictions}. If the actual outcome and the prediction differ, we count a logical error.
We summarize the simulation workflow in Figure\,\ref{fig:simulation_methods}. 

\subsection{Additional simulations}
\begin{figure}
    \centering
    \includegraphics[width=\linewidth]{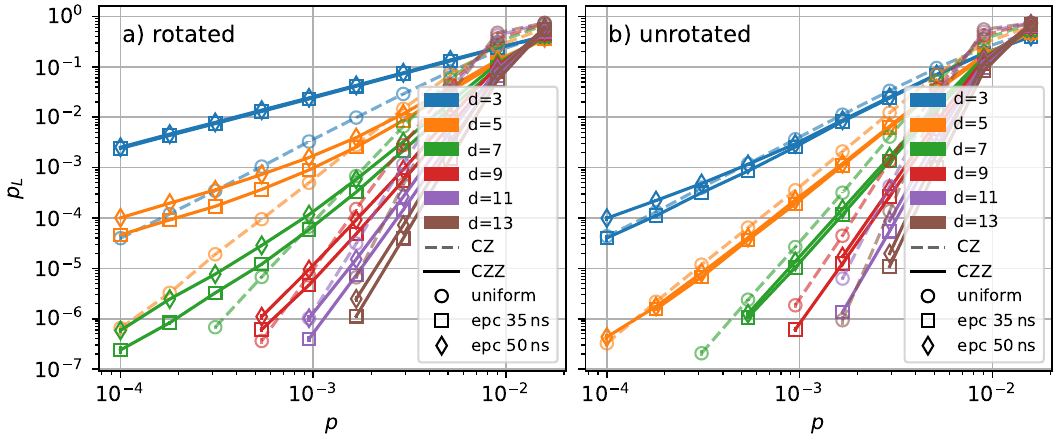}
    \caption{Logical error rates of the $50\,\mathrm{ns}$ gate compared to the the $35\,\mathrm{ns}$ gate and CZ circuit with uniform noise in Fig.\,\ref{fig:plot_50ns}. a) rotated and b) unrotated surface codes. The performance of the $50\,\mathrm{ns}$ gate is slightly worse, which can be explained by the smaller suppression of high degree Pauli marginals compared to the $35\,\mathrm{ns}$ gate. }
    \label{fig:plot_50ns}
\end{figure}

We show the logical error rate of the $50\,\mathrm{ns}$ gate compared to the the $35\,\mathrm{ns}$ gate and CZ circuit with uniform noise in Fig.\,\ref{fig:plot_50ns}. The performance of the $50\,\mathrm{ns}$ gate is slightly worse, which can be explained by the smaller suppression of high-degree Pauli marginals compared to the $35\,\mathrm{ns}$ gate, cf. Fig.\,\ref{fig:failing_paulis_rotated_5_5_True_SI1000_2}.

\subsection{Finite size scaling analysis of thresholds}
\begin{figure*}
    \centering
    \includegraphics[width=\linewidth]{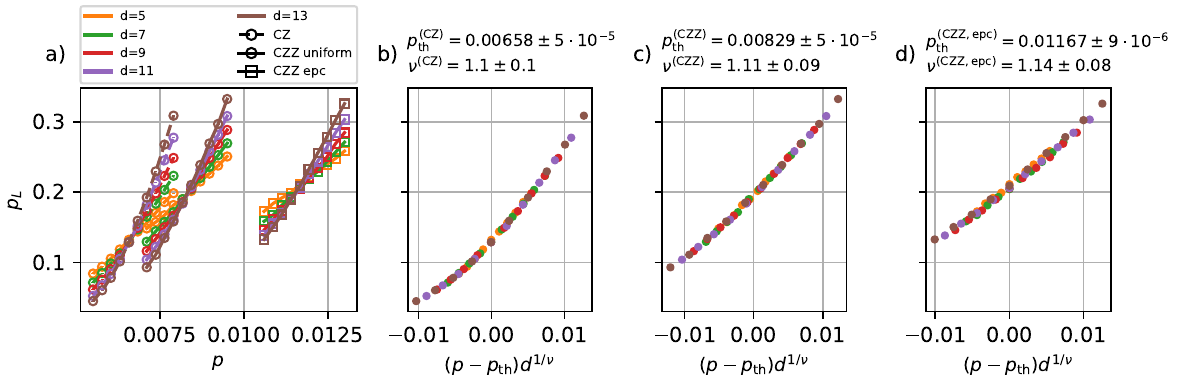}
    \includegraphics[width=\linewidth]{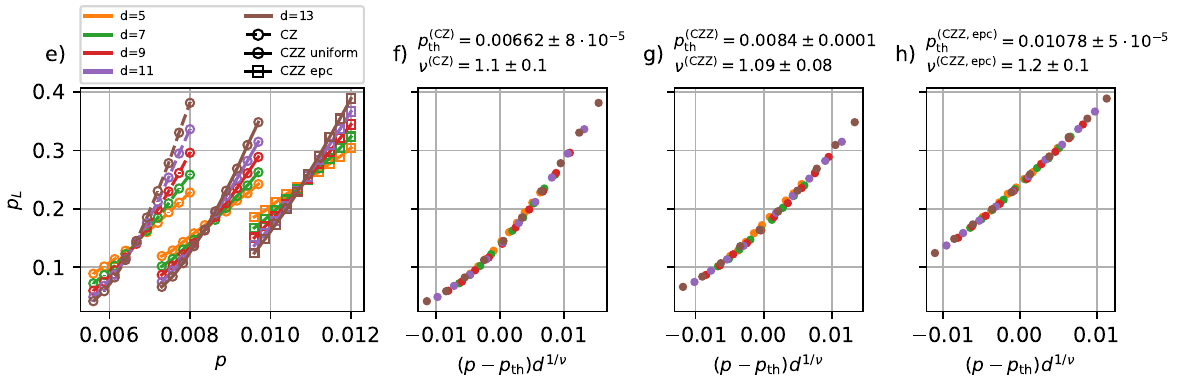}
    \caption{Threshold plots and finite size scaling analysis for circuits shown in the main text. a) - d) Rotated surface codes show an increase in threshold from $p_{\mathrm{th}}^{(\mathrm{CZ})} \approx 0.66\%$ to $p_{\mathrm{th}}^{(\mathrm{CZZ})} \approx 0.83\%$ and $p_{\mathrm{th}}^{(\mathrm{CZZ, epc})} \approx 1.2\%$. e) - h) Unrotated surface codes show an increase in threshold from $p_{\mathrm{th}}^{(\mathrm{CZ})} \approx 0.66\%$ to $p_{\mathrm{th}}^{(\mathrm{CZZ})} \approx 0.84\%$ and $p_{\mathrm{th}}^{(\mathrm{CZZ, epc})} \approx 1.1\%$.  }
    \label{fig:threshold_plots_fss}
\end{figure*}

We simulate the circuits of the main text with physical error rates in the vicinity of the threshold and perform a finite size scaling analysis using \texttt{pyfssa}\,\cite{Sorge2015}. We show the results in Fig.\,\ref{fig:threshold_plots_fss}. For both, rotated and unrotated surface codes, the threshold increases about $25\%$ for a CZZ gate implementation with a uniform depolarizing noise model and another $30-40\%$ with the effective Pauli channel. 

\end{document}